\documentclass[usenatbib]{mn2e} \usepackage{psfig}

\title[ULTRACAM photometry of the CVs GY~Cnc, IR~Com and HT~Cas]
  {ULTRACAM photometry of the eclipsing cataclysmic variables GY~Cnc,
  IR~Com and HT~Cas}

\author[W.\,J.\,Feline et al.] {W.\,J.\ Feline,$^1$\thanks{E-mail:
  w.feline@shef.ac.uk} V.\,S.\ Dhillon,$^1$ T.\,R.\ Marsh,$^{2}$
  C.\,A.\,Watson$^{1}$ and S.\,P.\ Littlefair$^1$\\ $^1$Department of
  Physics and Astronomy, University of Sheffield, Sheffield, S3 7RH,
  UK\\ $^2$Department of Physics, University of Warwick, Coventry CV4
  7AL, UK\\}

\date{\center{\Large Submitted for publication in the Monthly
    Notices of the Royal Astronomical Society}}

\begin{document}
  \maketitle

\begin{abstract}
We present high-speed, three-colour photometry of the eclipsing
cataclysmic variables GY~Cnc, IR~Com and HT~Cas. We find that the
sharp eclipses in GY~Cnc and IR~Com are due to eclipses of the white
dwarf. There is some evidence for a bright spot on the edge of the
accretion disc in GY~Cnc, but not in IR~Com. Eclipse mapping of HT~Cas
is presented which shows changes in the structure of the quiescent
accretion disc. Observations in 2002 show the accretion disc to be
invisible except for the presence of a bright spot at the disc
edge. 2003 observations, however, clearly show a bright inner disc and
the bright spot to be much fainter than in 2002. Although no outburst
was associated with either set of quiescent observations, the system
was $\sim0.6$~mJy brighter in 2003, mainly due to the enhanced
emission from the inner disc. We propose that these changes are due to
variations in the mass transfer rate from the secondary star and
through the disc. The disc colours indicate that it is optically thin
in both its inner and outer regions. We estimate the white dwarf
temperature of HT~Cas to be $15\,000\pm1000$~K in 2002 and
$14\,000\pm1000$~K in 2003.
\end{abstract}



\begin{table*}
\begin{center}
\caption{Journal of observations. The cycle number is determined
  from the ephemeris given in section \ref{sec:ephemerides}. Observing
  conditions were clear  except for 2002 September 14 and 2003 May 19
  and 21, when thin cirrus was present. The dead-time between
  exposures was 0.025~sec for all the observations. The {\em
  i}$^{\prime}$ sensitivity was lost during the eclipse of HT~Cas on
  2002 September 14 due to a technical problem with this band. The GPS
  signal, used for time-stamping each exposure, was lost for the HT~Cas
  data of 2003 October 29. This means that the absolute time of each
  exposure was incorrectly recorded, although the relative timing
  within the run remains accurate. The cycle number for the data of
  2003 October 29 is therefore estimated from times in the observing
  log. Due to poor weather, the extinction could not be measured for
  this night, and is therefore assumed to be 0.1~mag/airmass in the
  $r^{\prime}$ band, the mean of the previous and subsequent nights.}
\begin{tabular}{cccccccccc}
\hline Target & UT date & Cycle & Filters & Exposure &  No.\ of
 & No.\ of & Seeing \\
 & (start of night) & & & time (s) & frames & eclipses & (arcsec)\\

\hline GY~Cnc & 2003 May 19 & 6826 & {\em u}$^{\prime}${\em
g}$^{\prime}${\em z}$^{\prime}$ & 2.1 & 2256 & 1 & $>3$ \\

GY~Cnc & 2003 May 23 & 6849 & {\em u}$^{\prime}${\em g}$^{\prime}${\em
i}$^{\prime}$ & 1.6 & 2150 & 1 & 1.0 \\

IR~Com & 2003 May 21 & 37857 & {\em u}$^{\prime}${\em
g}$^{\prime}${\em i}$^{\prime}$ & 3.2 & 676 & 1 & 1.0 \\

IR~Com & 2003 May 23 & 37880 & {\em u}$^{\prime}${\em
g}$^{\prime}${\em i}$^{\prime}$ & 3.2 & 720 & 1 & 1.0 \\

IR~Com & 2003 May 25 & 37902 & {\em u}$^{\prime}${\em
g}$^{\prime}${\em i}$^{\prime}$ & 3.2 & 901 & 1 & 1.5 \\

HT~Cas & 2002 Sep.\ 13 & 119537 & {\em u}$^{\prime}${\em
g}$^{\prime}${\em i}$^{\prime}$ & 1.1 & 5651 & 1 & 1.2 \\

HT~Cas & 2002 Sep.\ 14 & 119550 & {\em u}$^{\prime}${\em
  g}$^{\prime}${\em i}$^{\prime}$ & 0.97--1.1 & 5470 & 1 & 1.3--2.3 \\

HT~Cas & 2003 Oct.\ 29 & 125116 & {\em u}$^{\prime}${\em
g}$^{\prime}${\em i}$^{\prime}$ & 1.3 & 4659 & 1 & 1.4 \\

HT~Cas & 2003 Oct.\ 30 & 125129--30 & {\em u}$^{\prime}${\em
g}$^{\prime}${\em i}$^{\prime}$ & 1.3 & 6930 & 2 & 1.0--1.5 \\

\hline
\end{tabular}
\label{tab:journal}
\end{center}
\end{table*}

\section{Introduction}
\label{sec:introduction}

Cataclysmic variable stars (CVs) are a type of close binary system in
which the white dwarf primary star (of mass $M_{{\rm w}}$) accretes
matter from the secondary  star (of mass $M_{{\rm r}}$), usually a cool
main-sequence star. This accretion occurs via a ballistic gas stream
originating from the inner Lagrangian (L1) point and an
accretion  disc surrounding the primary. A `bright spot' is frequently
formed at the point of impact of the gas stream and accretion
disc. Foreshortening of this feature gives rise to an orbital hump in
the light curve as the bright spot rotates into view.

If the rate of mass transfer into the disc is greater than the rate of
flow of material through the disc, matter will build up in the disc
until a critical density is reached, whereupon a thermal instability
\citep{osaki74,meyer81,smak82,smak84,faulkner83,mineshige83} causes
the disc viscosity to greatly increase, resulting in a huge rise
in the rate of mass transfer through the disc. The result is an
increase in the system brightness of between two and five magnitudes,
known as a dwarf nova eruption, the presence of which defines the
sub-class of CVs known as the dwarf novae. This disc instability model
has been reviewed several times \citep[e.g.][]{osaki96,lasota01}. For
an excellent review of CVs in general, see \citet{warner95}.

\defcitealias{feline04a}{Feline et al.\ 2004a}
\defcitealias{feline04b}{b}
\defcitealias{feline04c}{c}
\defcitealias{kato02b}{Kato, Ishioka \& Uemura (2002b)}
\defcitealias{kato02a}{Kato, Baba \& Nogami 2002a}

The light curves of eclipsing CVs reveal a wealth of information about
these objects. Successive eclipses of the white dwarf, bright spot and
accretion disc by the secondary star can be used to determine the
system parameters to a high degree of accuracy (see, for example,
\citetalias{feline04a},\citetalias{feline04b},\citetalias{feline04c}).
Spatial structure in the accretion disc can be directly determined
from observations using the  eclipse mapping technique developed by
\citet{horne85} and recently reviewed by \citet{baptista01a}.

GY~Cnc (= RX~J0909.8+1849 = HS~0907+1902) is a $V\sim16$ eclipsing
dwarf nova with an orbital period  $P_{{\rm orb}}=4.2$~hr. GY~Cnc was
detected in both the Hamburg Schmidt objective prism survey
\citep{hagen95} and the ROSAT Bright Source catalogue \citep{voges99},
and identified as a possible CV by \citet{bade98}. Spectroscopic and
photometric follow-up observations by \citet{gansicke00} confirmed the
status of GY~Cnc as an eclipsing dwarf nova by detecting it in both
outburst and quiescence. \citet{shafter00} used multi-colour
photometric observations of GY~Cnc to determine the temperatures of
the white dwarf, bright spot and accretion disc and the disc power-law
temperature exponent, which they found to be largely independent of
the mass ratio assumed. Spectroscopic and photometric observations
obtained by \citet{thorstensen00} constrain the mass ratio
$q=M_{{\rm r}}/M_{{\rm w}}=0.41\pm0.04$ and the orbital inclination
$i=77\fdg0\pm0\fdg9$ (after applying corrections to the radial
velocity of the secondary star $K_{2}$). The spectral type of the
secondary star has been estimated  as M$3\pm1.5$
\citep{gansicke00,thorstensen00}. GY~Cnc was observed during decline
from outburst in 2001 November by \citetalias{kato02b}, who suggest that
GY~Cnc is an ``above-the-gap counterpart'' to the dwarf nova HT~Cas.

IR~Com (= S~10932~Com) was discovered as the optical counterpart to
the ROSAT X-ray source RX~J1239.5 \citep{richter95}. IR~Com exhibits
high (16.5~mag) and low (18.5~mag) brightness states
\citep{richter95,richter97}, with outburst amplitudes of $\sim4.5$~mag
(\citealt{richter97}; \citetalias{kato02a}). \citet*{wenzel95}
detected eclipses in the light curve of IR~Com and determined an
orbital period of 2.1~hr, just below the period gap. \citet{richter97}
present photometric and spectroscopic observations of IR~Com, which
illustrate the highly variable nature of the
target. \defcitealias{kato02a}{Kato et al.\ (2002a)}
\citetalias{kato02a} reported observations of IR~Com at, and during,
the decline from outburst. None of the published light curves of
IR~Com show much evidence for the presence of an orbital hump in the
light curve before eclipse, or for asymmetry of the eclipse itself
(although the limited time-resolution of the observations may mask
such asymmetries to an extent). \citetalias{kato02a} again suggest
that IR~Com is a twin of HT~Cas.

HT~Cas is a well-known and well-studied faint eclipsing dwarf nova. It
has a quiescent magnitude of $V\sim16.4$ and an orbital period of
1.77~hr. The literature on HT~Cas is extensive; here we only discuss a
selection of relevant work. The system parameters of HT~Cas have been
well-determined by \citet*{horne91b} using simultaneous {\em U, B, V}
and {\em R} observations in conjunction with those of
\citet{patterson81}: $q=0.15\pm0.03$ and $i=81\fdg0\pm1\fdg0$. In a
companion paper, \citet*{wood92} determined the temperature of the
white dwarf ($T=14\,000\pm1000$~K) and estimated the distance to the
system ($D=125\pm8$~pc). They also eclipse-mapped the accretion disc,
illustrating the flat radial temperature profile typical of quiescent
dwarf novae. \citet*{vrielmann02} have recently reconstructed the
temperatures and surface densities of the quiescent accretion disc of
HT~Cas using physical parameter eclipse mapping. This method also
yields an estimate of the distance, $D=207\pm10$~pc. \citet{marsh90c}
detected the secondary star in HT~Cas using low-resolution spectra,
estimating the spectral type as M$5.4\pm0.3$. \citet{marsh90c} found
the secondary star to be consistent with main-sequence values for the
mass, radius and luminosity. \citet{robertson96} discuss the long-term
quiescent light curve of HT~Cas, with particular regard to the
(unusual) presence of high- and low-states (at 16.4 and
17.7~mag). \citet{wood95a} detected an X-ray eclipse of HT~Cas using
ROSAT observations during one of the system's low-luminosity
states. The X-rays are believed to originate in a boundary layer
between the white dwarf and inner accretion disc. 

In this paper, we present light curves of GY~Cnc, IR~Com and HT~Cas,
obtained with ULTRACAM, an ultra-fast, triple-beam CCD camera; for
more details see \citet{dhillon01b} and Dhillon et al.\ (in
preparation). Our data for GY~Cnc and IR~Com are of the highest
time-resolution yet obtained, and are the first simultaneous,
three-colour light curves for these objects. We present eclipse maps
of HT~Cas in quiescence in 2002 and 2003, which show distinct changes
in the structure of the accretion disc which are related to the
overall brightness of the system.

\begin{figure*}
\begin{tabular}{ccc}
\psfig{figure=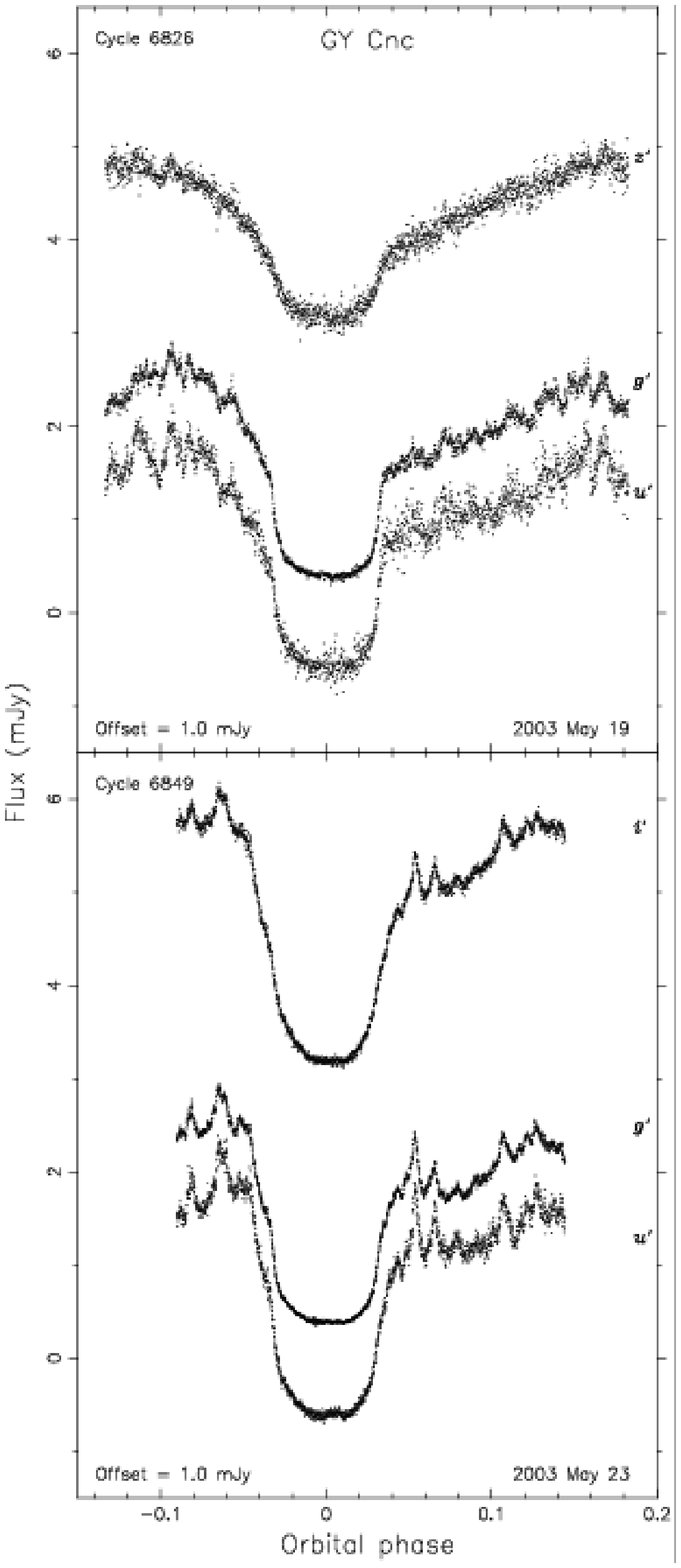,width=5.5cm,angle=0.} &
\psfig{figure=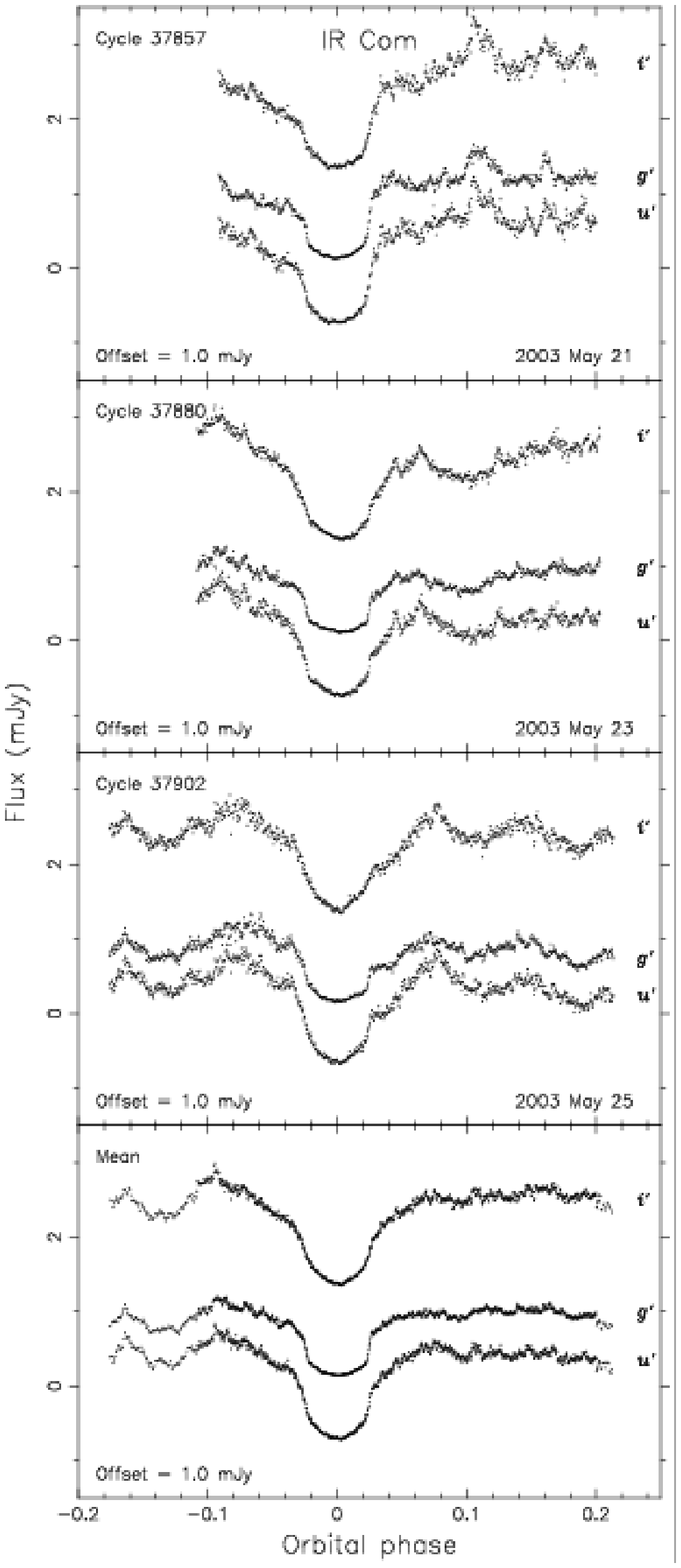,width=5.5cm,angle=0.} &
\psfig{figure=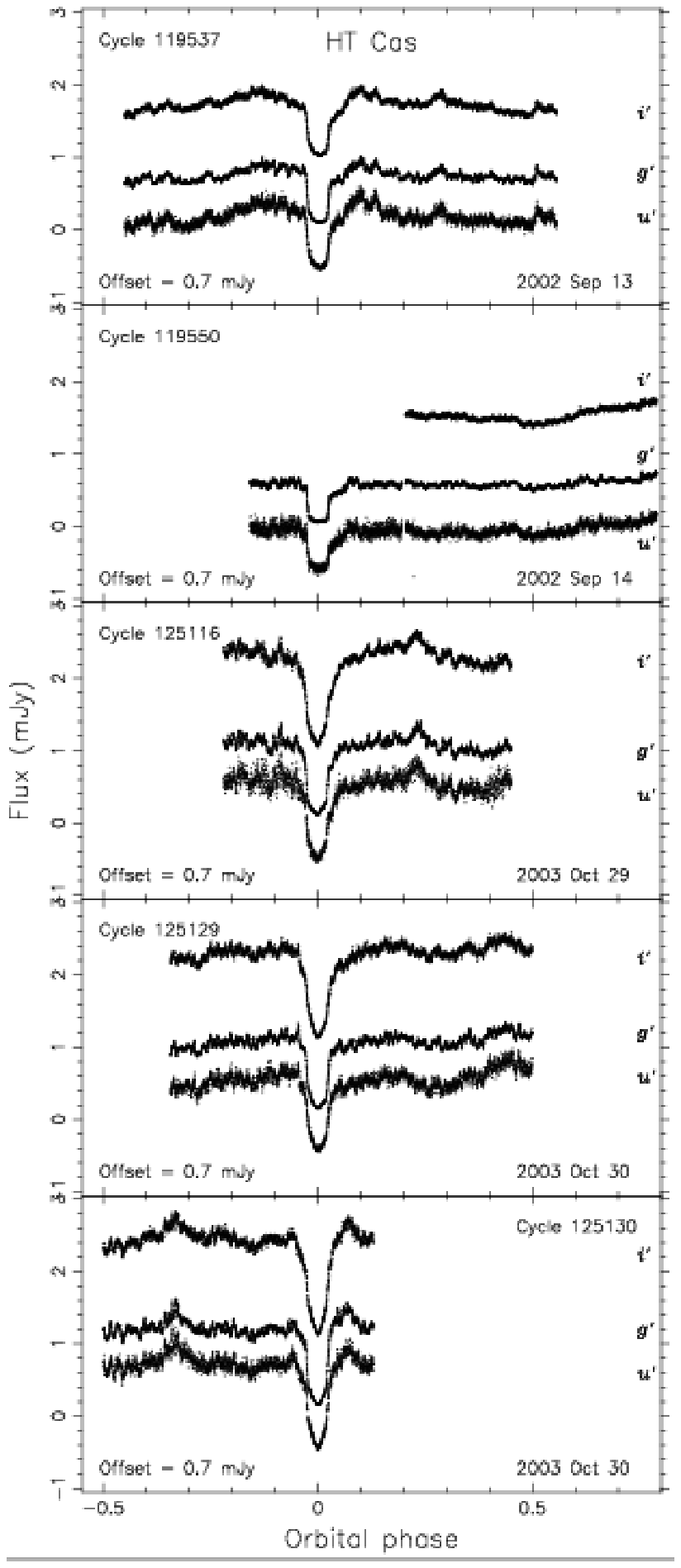,width=5.5cm,angle=0.} 
\end{tabular}
\caption{Left. The light curves of GY~Cnc. Centre. The light curves of
  IR~Com. Right. The light curves of HT~Cas. The {\em i}$^{\prime}$ and
  {\em z}$^{\prime}$ data are offset vertically upwards and the {\em
  u}$^{\prime}$ data are offset vertically downwards by the amount
  specified in the relevant plot. Note that the {\em i}$^{\prime}$
  HT~Cas data of 2002 September 14 were lost due to a technical
  problem with this CCD. The mean light curve of IR~Com is also shown.}
\label{fig:lightcurves}
\end{figure*}


\section{Observations}
\defcitealias{feline04b}{Feline et al.\ (2004b}

GY~Cnc, IR~Com and HT~Cas were observed using ULTRACAM
(\citealt{dhillon01b}; Dhillon et al., in preparation) on the 4.2-m
William Herschel Telescope (WHT) at the Isaac Newton Group of
Telescopes, La Palma. The data were obtained in three colour bands
simultaneously. The observations are summarised in
Table~\ref{tab:journal}. Data reduction was carried out as described
in \citetalias{feline04b},\citetalias{feline04c}) using the ULTRACAM
pipeline data reduction software. All HJD times quoted are UTC
corrected to the heliocentre (i.e.\ not TDB). The resulting light
curves of GY~Cnc, IR~Com and HT~Cas are shown in
Fig.~\ref{fig:lightcurves}.


\section{Orbital ephemerides}
\label{sec:ephemerides}

The times of white dwarf mid-ingress $T_{{\rm wi}}$ and mid-egress
$T_{{\rm we}}$ were determined by locating the times when the minimum
and maximum values, respectively, of the light curve derivative
occurred \citep*{wood85}. The times of mid-eclipse $T_{{\rm mid}}$
given in Table~\ref{tab:eclipse_times} were determined by assuming the
white dwarf eclipse to be symmetric around phase zero and taking
$T_{{\rm mid}}=(T_{{\rm we}}+T_{{\rm wi}})/2$. If the sharp eclipse is
caused by the obscuration of the bright spot rather than the white
dwarf, then phase zero, as defined by the ephemerides below, may not
necessarily correspond to the conjunction of the white and red dwarf
components. As discussed in sections~\ref{sec:gycnc}--\ref{sec:htcas},
however, it is probable that the sharp eclipse in all three objects
(and certainly HT~Cas) is of the white dwarf.

\defcitealias{kato02b}{Kato et al.\ (2002b)}

The orbital ephemeris of GY~Cnc was determined using the seven eclipse
timings of \citet{gansicke00}, the eight timings of \citet{shafter00},
the seven timings of \citet{kato00}, the two timings of
\citet{vanmunster4210}, the four timings of \citetalias{kato02b} and
the six ULTRACAM timings determined in this paper and given in
Table~\ref{tab:eclipse_times}. Errors adopted were
$1\times10^{-4}$~days for the data of \citet{gansicke00} and
\citet{shafter00}, $5\times10^{-5}$~days for the data of
\citet{kato00}, \citet{vanmunster4210} and \citetalias{kato02b} and
$1\times10^{-5}$~days for the ULTRACAM data. A linear least squares
fit to these times gives the following orbital ephemeris for GY~Cnc:
\begin{displaymath}
\begin{array}{ccrcrl}
\\ HJD & = & 2451581.826653 & + & 0.1754424988 & E.  \\
 & & 14 & \pm & 21 &
\end{array} 
\end{displaymath}

\begin{table*}
\begin{center}
\caption{Mid-eclipse timings. The cycle numbers were determined from the
  ephemerides described in section~\ref{sec:ephemerides}. Note that a
  technical problem with the {\em i}$^{\prime}$ CCD corrupted the
  data during eclipse in this band on 2002 September 14.}
\begin{tabular}{lccccccc}
\hline
Target & UT Date & Cycle & \multicolumn{3}{c}{${\rm HJD}+2\,452\,530$}\\
 & (start of night) & & {\em u}$^{\prime}$ & {\em g}$^{\prime}$ & {\em
  i}$^{\prime}$ & {\em z}$^{\prime}$\\
\hline
GY~Cnc & 2003 May 19 & 6826 & 249.397235 & 249.397223 & -- &
249.397248 \\
GY~Cnc & 2003 May 23 & 6849 & 253.432215 & 253.432254 &
253.432254 & -- \\

IR~Com & 2003 May 21 & 37857 & 251.503269 & 251.503250 &
251.503194 & -- \\ 
IR~Com & 2003 May 23 & 37880 & 253.505096 & 253.505153 &
253.505153 & -- \\
IR~Com & 2003 May 25 & 37902 & 255.419890 & 255.419966 &
255.419909 & -- \\ 

HT~Cas & 2002 Sep.\ 13 & 119537 & 1.503015 & 1.503035 & 1.502995 & -- \\
HT~Cas & 2002 Sep.\ 14 & 119550 & 2.460443 & 2.460477 & -- & -- \\
HT~Cas & 2003 Oct.\ 30 & 125129 & 413.338089 & 413.338159 & 413.338199 &
-- \\ 
HT~Cas & 2003 Oct.\ 30 & 125130 & 413.411832 & 413.411792 & 413.411769 &
-- \\ 
\hline
\end{tabular}
\label{tab:eclipse_times}
\end{center}
\end{table*}

\defcitealias{kato02a}{Kato et al.\ 2002a}

To determine the orbital ephemeris of IR~Com, we used the 24 timings
of \citet[as listed in \citetalias{kato02a}]{richter97},
\defcitealias{kato02a}{Kato et al.\ (2002a)} the 14 eclipse timings of
\citetalias{kato02a} and those nine, given in
Table~\ref{tab:eclipse_times}, determined from the ULTRACAM data. The
errors adopted for the data of \citet{richter97} and
\citetalias{kato02a} were $\pm1\times10^{-3}$~days for cycles
$-134516$, $-51035$, $-42189$, $-29597$ and $-21531$ and
$\pm5\times10^{-5}$~days for subsequent cycles, except where stated
otherwise by \citetalias{kato02a}.  Those adopted for the ULTRACAM
timings were $\pm2\times10^{-5}$~days. The orbital ephemeris of IR~Com
was determined by a linear least squares fit to the above timings, and
is
\begin{displaymath}
\begin{array}{ccrcrl}
\\ HJD & = & 2449486.4818691 & + & 0.08703862787 & E.  \\
 & & 26 & \pm & 20 &
\end{array} 
\end{displaymath}

To determine the orbital ephemeris of HT~Cas we used the 11
mid-eclipse times of \citet{patterson81}, the 23 times of
\citet*{zhang86}, the 15 times of \citet{horne91b} and the 11 ULTRACAM
times given in Table~\ref{tab:eclipse_times}. The times of
\citet{patterson81}, \citet{zhang86} and \citet{horne91b} were
assigned errors of $5\times10^{-5}$~days and the times in
Table~\ref{tab:eclipse_times} assigned errors of
$5\times10^{-6}$~days. A linear least squares fit to these times gives
the following orbital ephemeris for HT~Cas:
\begin{displaymath}
\begin{array}{ccrcrl}
\\ HJD & = & 2443727.937290 & + & 0.07364720309 & E.\\ & & 8 & \pm
 & 7 &
\end{array} 
\end{displaymath}

The loss of accurate timings for the 2003 October 29 HT~Cas data meant
that these data were phased according to the orbital period derived
above, with the mid-point of the observed eclipse as the
zero-point. The cycle number was accurately determined from the times
in the hand-written observing log. This may result in a slight fixed
time offset for these data due to the uncertainty in determining the
point of mid-eclipse.

These ephemerides were used to phase all of our data.


\section{GY~Cnc}
\label{sec:gycnc}

In keeping with previous observations (summarised in
section~\ref{sec:introduction}), the light curve of GY~Cnc shown in
Fig.~\ref{fig:lightcurves} shows a deep primary eclipse, with the {\em
g}$^{\prime}$ flux dropping from a peak value of approximately 3~mJy
(15.2~mag) to about 0.6~mJy (17.0~mag) at mid-eclipse. This places the
system slightly above its quiescent brightness of $V=16$, shortly
after an outburst which reached twelfth-magnitude on 2003 May 13
(Waagen, private communication; observed by the amateur organisation
the American Association  of Variable Star Observers, AAVSO). The
system was therefore likely to still be in decline from outburst. The
eclipse morphology appears to be that of a gradual disc eclipse with a
sharp eclipse of the white dwarf or bright spot superimposed
thereon. We suspect that the sharp eclipse is that of a white dwarf,
not the bright spot, as in both cycles observed the ingress and egress
are of the same order in terms of both duration and depth. The eclipse
is flat-bottomed, suggesting that the disc and white dwarf are
completely obscured at these phases. The eclipse of the disc appears
to be asymmetric, with the ingress being rather sharper than the more
gradual egress. This is indicative of asymmetry in the disc structure,
possibly due to an extended bright spot at the disc rim. The changing
foreshortening of the bright spot, the cause of the orbital hump often
observed in other dwarf novae, would also account for the rather
greater flux before eclipse than after. Indeed, the ingress observed
on 2003~May~23 appears to show two steps, which we attribute to first
the bright-spot then the white dwarf entering eclipse. The likely
presence of an extended bright spot is another reason why we suspect
that the sharp, discrete eclipse visible in both nights' data is that
of the white dwarf. The light curve of Fig.~\ref{fig:lightcurves} is
morphologically similar to quiescent light curves in the literature
\citep{gansicke00,shafter00,thorstensen00}.

\section{IR~Com}
\label{sec:ircom}

The light curve of IR~Com, shown in Fig.~\ref{fig:lightcurves}, also
exhibits a deep primary eclipse. The light curve is highly variable
outside of eclipse, with a maximum {\em g}$^{\prime}$ flux of about
1.6~mJy (15.9~mag) and a minimum during eclipse of approximately
0.16~mJy (18.4~mag). The average out-of-eclipse {\em g}$^{\prime}$
flux level of IR~Com during our observations was 1.0~mJy (16.4~mag),
consistent with the system being in quiescence. From the light curve
of IR~Com shown in Fig.~\ref{fig:lightcurves} it is clear that the
eclipse morphology of this object is highly variable. There is a clear
eclipse of a compact structure, either the white dwarf or bright spot,
as evidenced by the sharpness of the ingress and egress. The highly
variable nature of the light curve of IR~Com makes it difficult to
determine whether the sharp eclipse is of the white dwarf or bright
spot. The mean light curve of IR~Com shown in
Fig.~\ref{fig:lightcurves} shows the main features of the light curve
much more clearly, as flickering is much reduced. The sharp eclipse is
revealed to be nearly symmetric, with evidence for an eclipse of the
disc in the V-shaped eclipse bottom and the slopes before and after
the sharp eclipse. No sign of the eclipse of another compact object is
seen, so the sharp eclipse must be of the white dwarf (in which case
the bright spot is extremely faint) or an eclipse of the bright spot
(in which case the white dwarf remains visible at all phases).

Contact phases of the sharp eclipse of IR~Com were determined using
the derivative of the light curve, as described by
\citetalias{feline04b},\citetalias{feline04c}, and references
therein). For reasons of space, we do not reproduce these contact
phases here, but we note that they do not show any evidence for
asymmetry in the duration of ingress and egress (as is frequently the
case with the eclipse of a bright spot, where the ingress is of a
longer duration than the egress). Additionally, using the
\citet{nauenberg72} mass--radius relation for a cold, rotating white
dwarf with Kepler's third law \defcitealias{feline04b}{Feline et al.\
2004b} (as described in \citetalias{feline04b},\citetalias{feline04c})
for reasonable values of $q$ shows that the eclipse contact phases are
entirely consistent with the eclipsed object being of the correct size
for a white dwarf. As \citetalias{kato02a} point out, their mid-eclipse
timings show no significant differences between outburst and
quiescence, implying that in both quiescence and outburst the
brightness distribution is centred on the white dwarf. These points
lead us to believe that the primary eclipse is of the white dwarf,
rather than of the bright spot.

No unambiguous bright spot feature is visible in either the individual
or mean light curves of IR~Com shown in
Fig.~\ref{fig:lightcurves}. From the absence of flickering during
primary eclipse, it  appears that the flickering is confined to the
inner regions of the accretion disc or the white dwarf itself. The
origin of the flickering in IR~Com is most likely the boundary layer
between the white dwarf and accretion disc.

\begin{figure}
\centerline{\psfig{figure=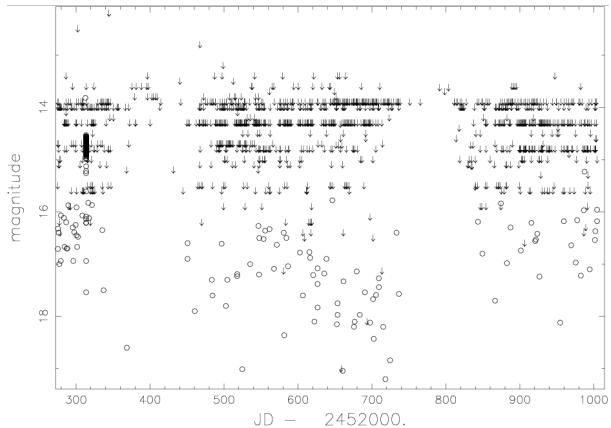,width=8.0cm,angle=0.}}
\caption{The long-term light curve of HT~Cas, courtesy of the AAVSO
  (Waagen, private communication). Open circles are V band
  observations; arrows mark upper limits on the magnitude of the
  system. The Julian date scale corresponds to calendar dates from
  2002 January 1 to 2004 January 1. Note the outburst in 2002
  February, which peaked on 2002 February 6 ($={\rm
  JD}\;2\,452\,312$).}
\label{fig:htcas_aavso}
\end{figure}

\begin{figure*}
\begin{tabular}{cc}
\psfig{figure=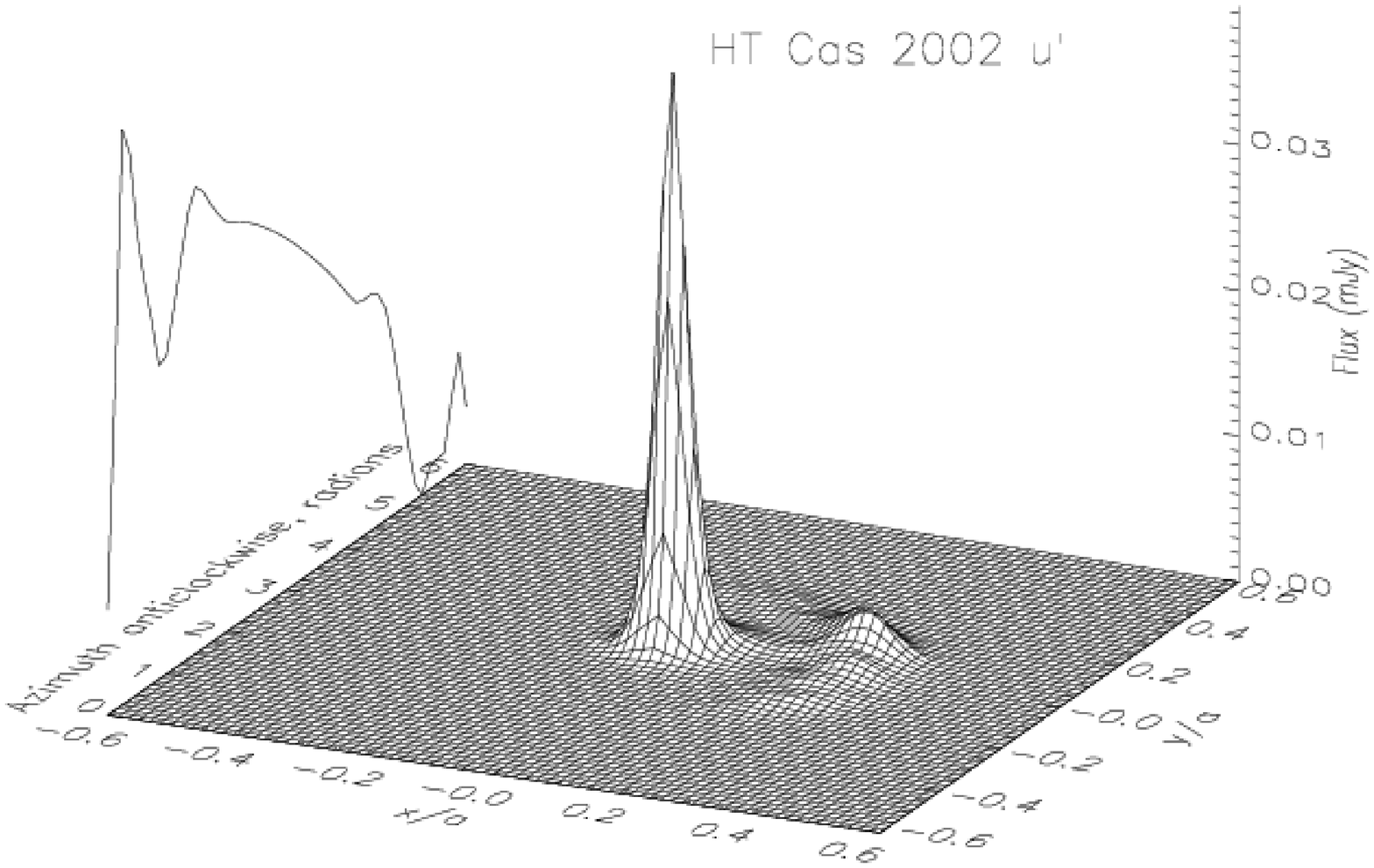,width=7.0cm,angle=0.} &
\psfig{figure=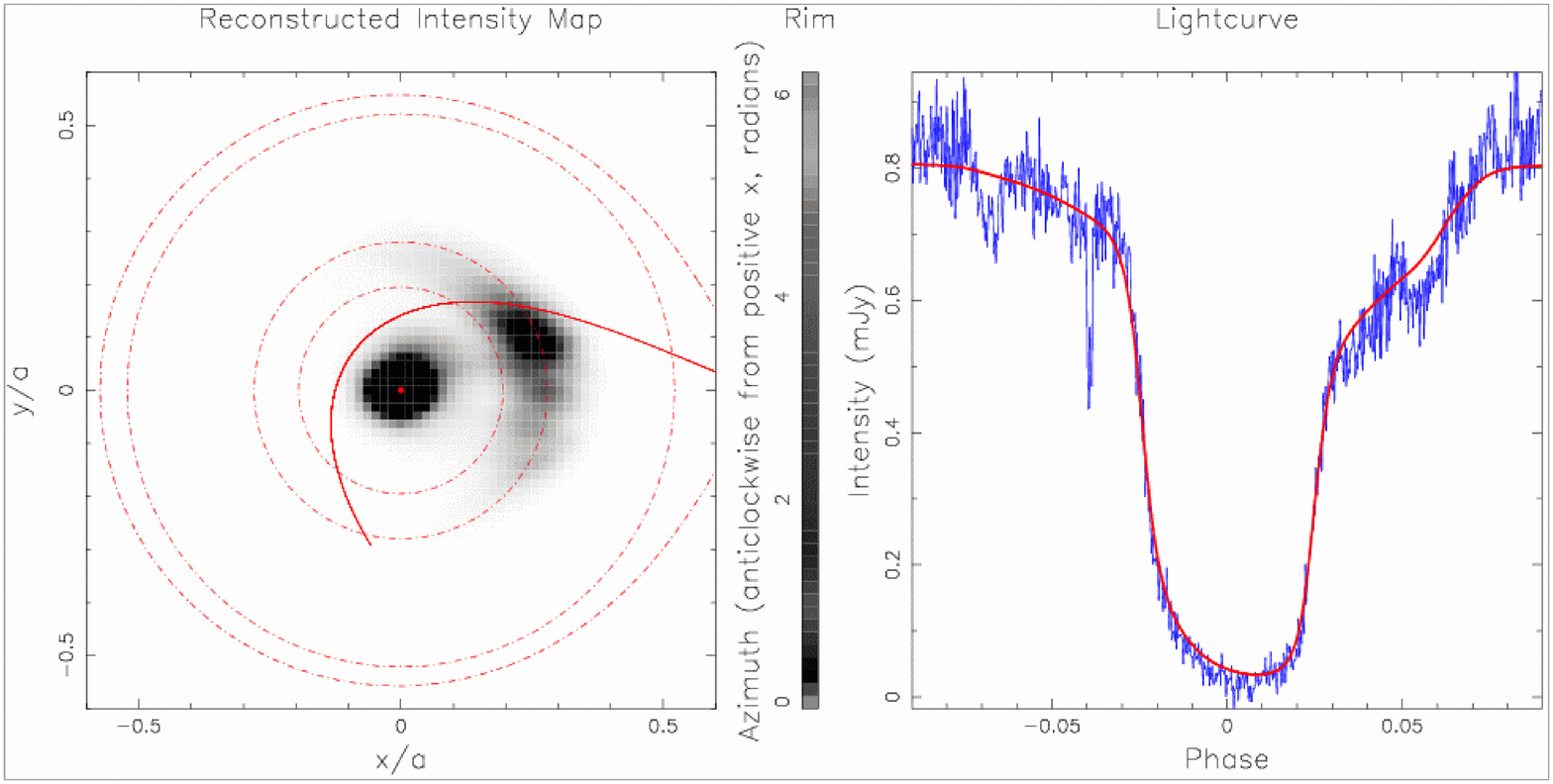,width=10.0cm,angle=0.} \\
\psfig{figure=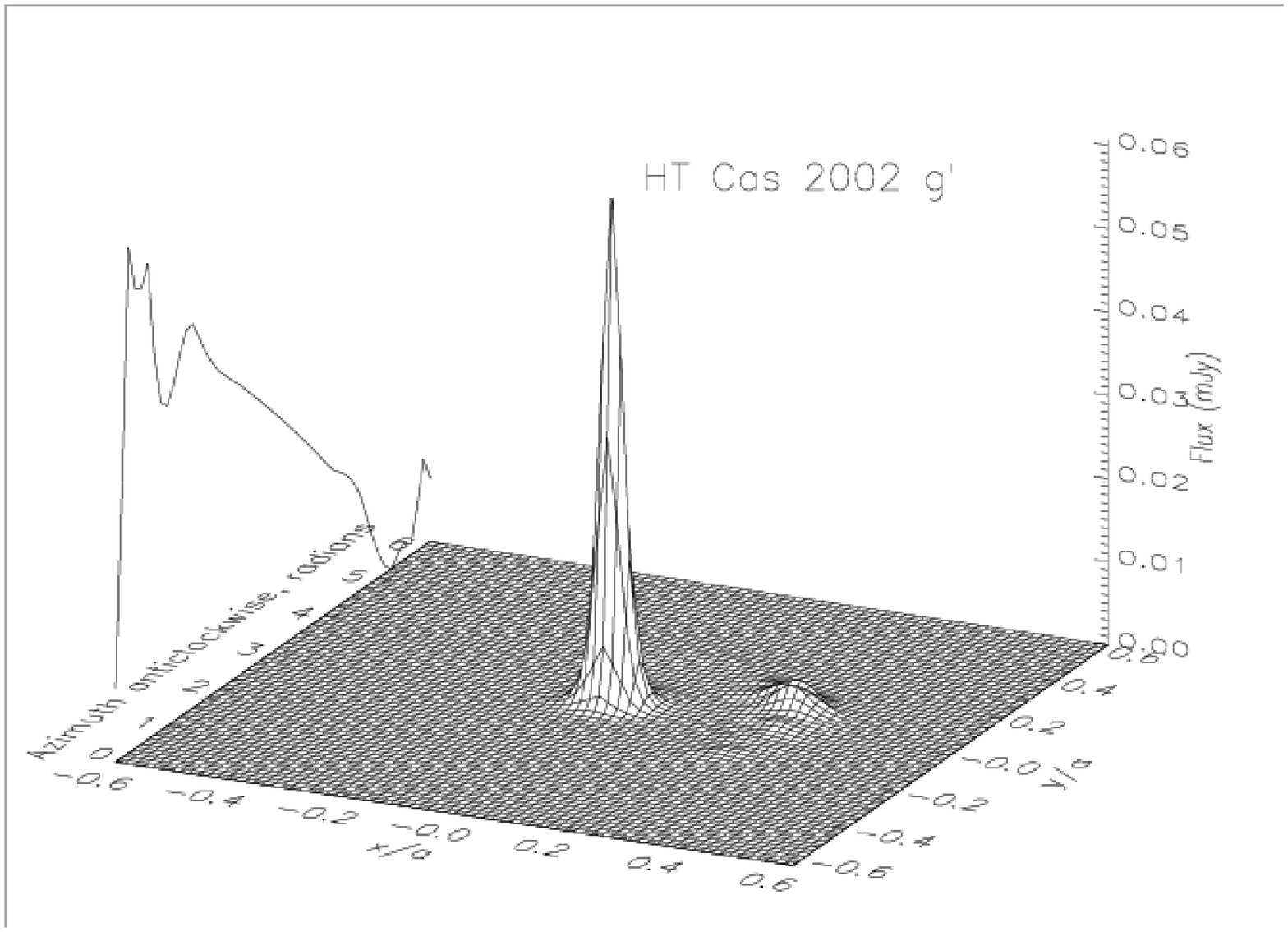,width=7.0cm,angle=0.} &
\psfig{figure=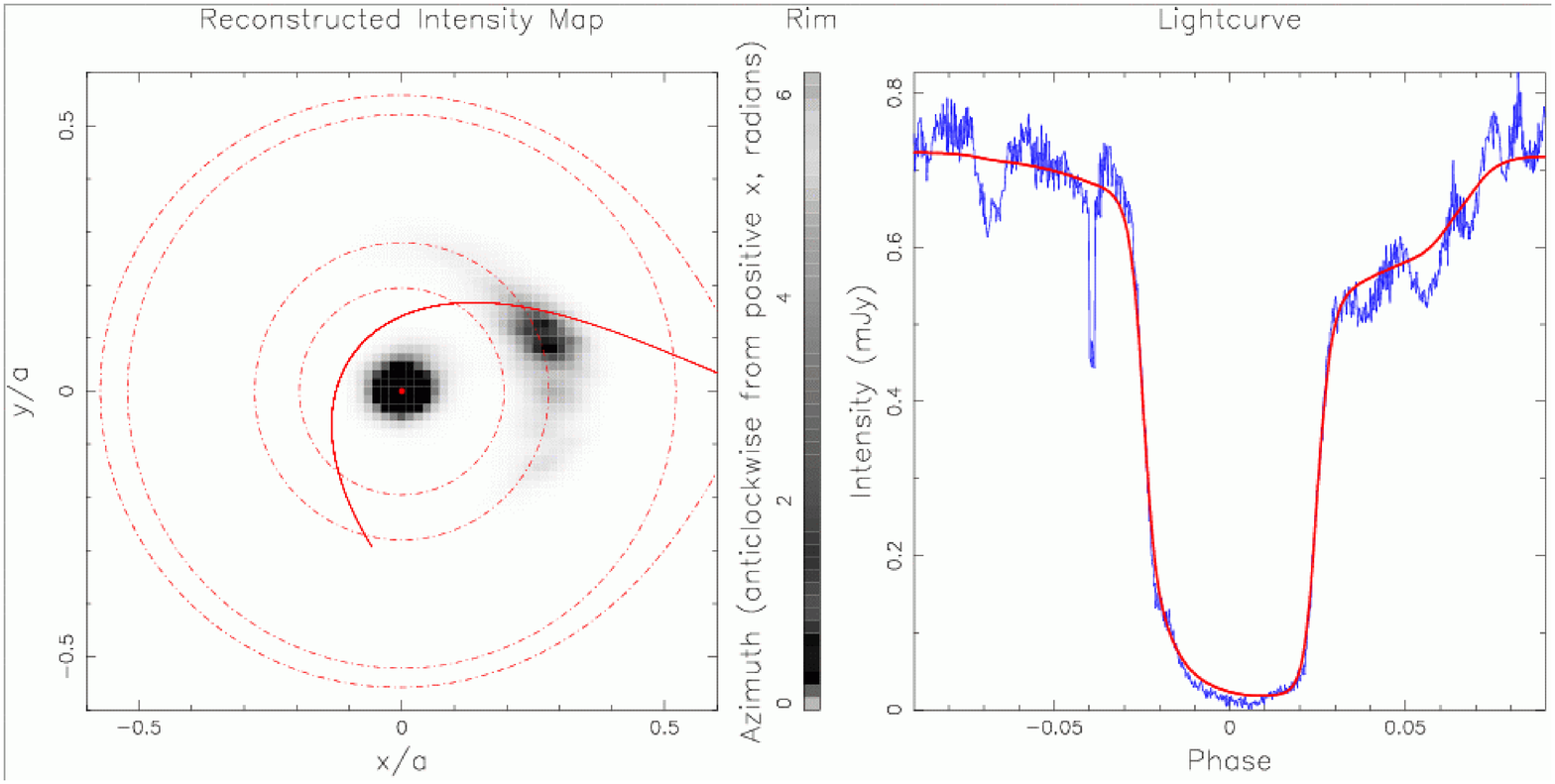,width=10.0cm,angle=0.} \\
\psfig{figure=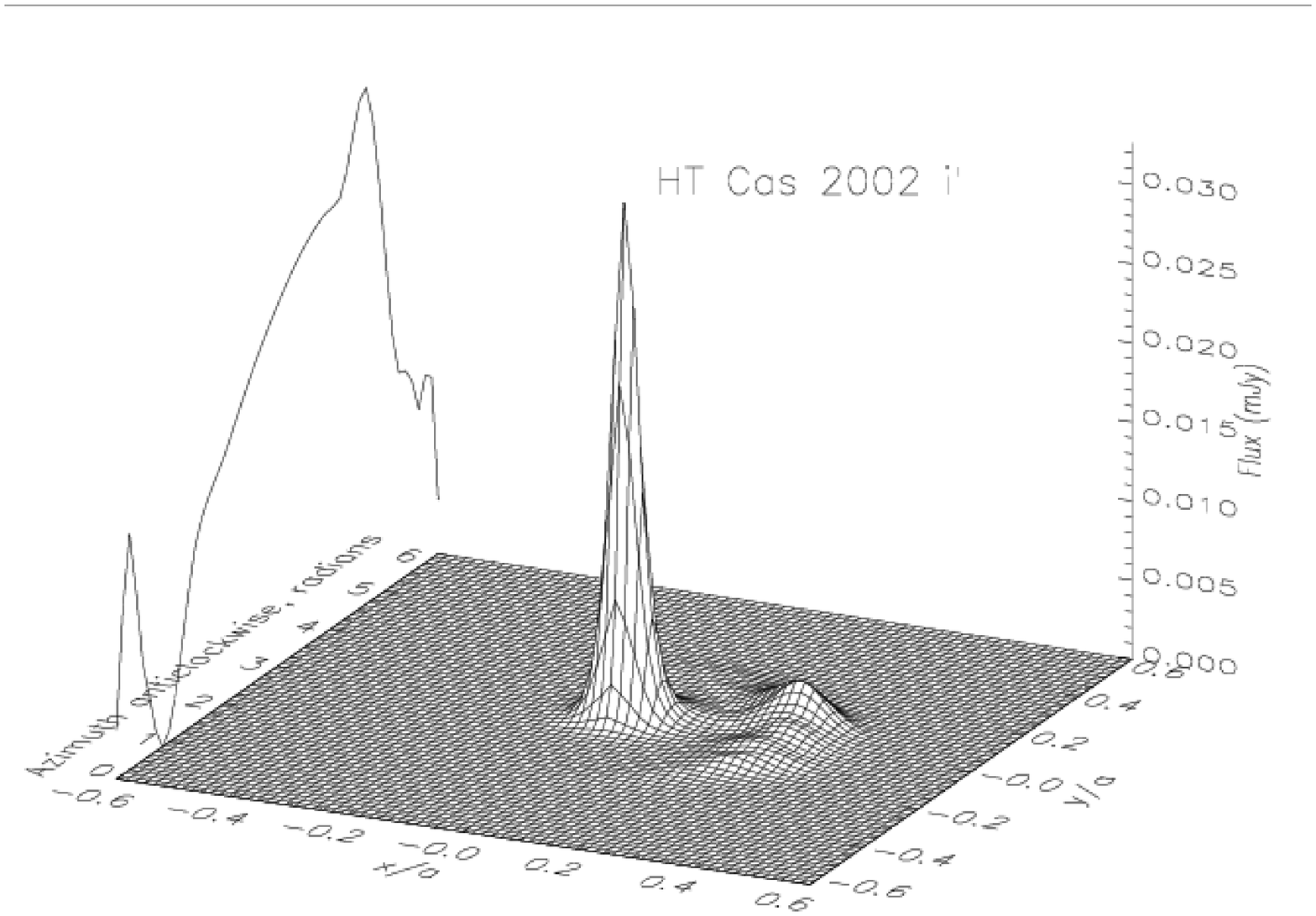,width=7.0cm,angle=0.} &
\psfig{figure=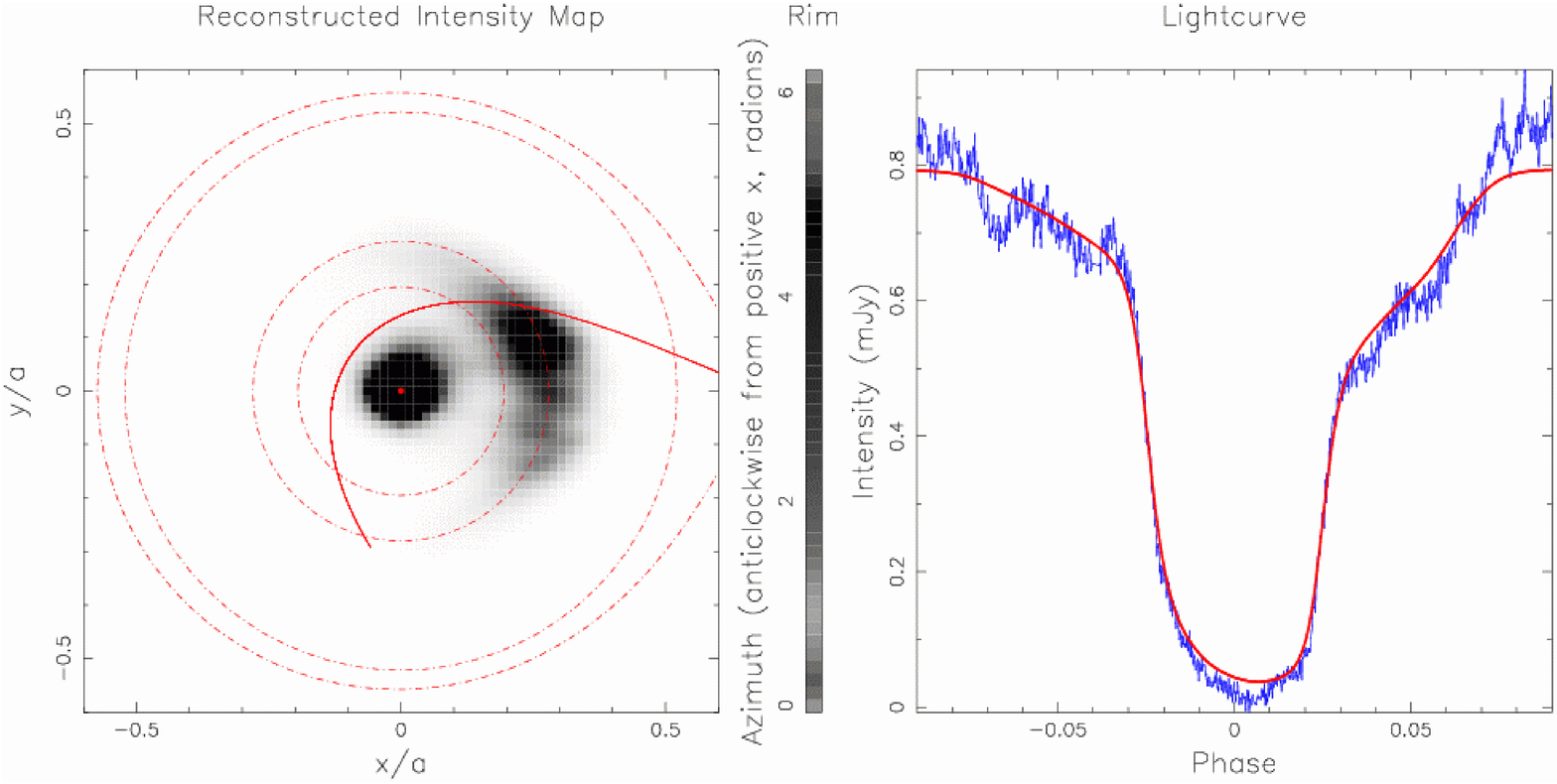,width=10.0cm,angle=0.} \\
\end{tabular}
\caption{Eclipse mapping of the 2002 September 13--14 accretion disc
  of HT~Cas. Top row. Left: a three-dimensional representation of the
  brightness distribution of the accretion disc of HT~Cas for the {\em
  u}$^{\prime}$ data of 2002 September 13--14. The white dwarf is at
  the origin, and the red dwarf at $(x,y)=(1,0)$. The rim intensity,
  as described in the text, is shown on the edge of the grid. Centre:
  a two-dimensional view of the brightness distribution of the
  accretion disc. The dot-dashed red lines are, from the centre out,
  the circularisation radius \citep[$0.1949a$;][ their equation
  13]{verbunt88}, the disc radius (the same radius as the rim;
  estimated from the radial position of the bright spot in the eclipse
  maps as $0.28a$), the tidal radius \citep[$0.5217a$;][]{paczynski77}
  and the primary star's Roche lobe. The solid red line is the
  trajectory of the gas stream. The rim intensity is shown to the
  right. The scale is linear from zero to five per cent of the
  maximum, with areas with intensities greater than 5 per cent of the
  maximum shown in black. Dark regions are brighter. Right: The light
  curve (blue) with the fit (red) corresponding to the intensity
  distribution shown on the left. The middle row shows the {\em
  g}$^{\prime}$ data for 2002 September 13--14. The bottom row shows
  the {\em i}$^{\prime}$ data for 2002 September 13 only, due to a
  problem with the red CCD on 2002 September 14. Prior to fitting, a
  (constant) offset was subtracted from the light curves. This offset
  was 0.15, 0.09 and 0.32~mJy for the {\em u}$^{\prime}$, {\em
  g}$^{\prime}$ and {\em i}$^{\prime}$ data, respectively. The sharp
  dip visible at about phase $-0.04$ in the {\em u}$^{\prime}$ and
  {\em g}$^{\prime}$ light curves is due to a short gap between
  observing runs on 2003 September 13 coinciding with a dip in the
  2003 September 14 light curve.}
\label{fig:htcas_2002}
\end{figure*}

\begin{figure*}
\begin{tabular}{cc}
\psfig{figure=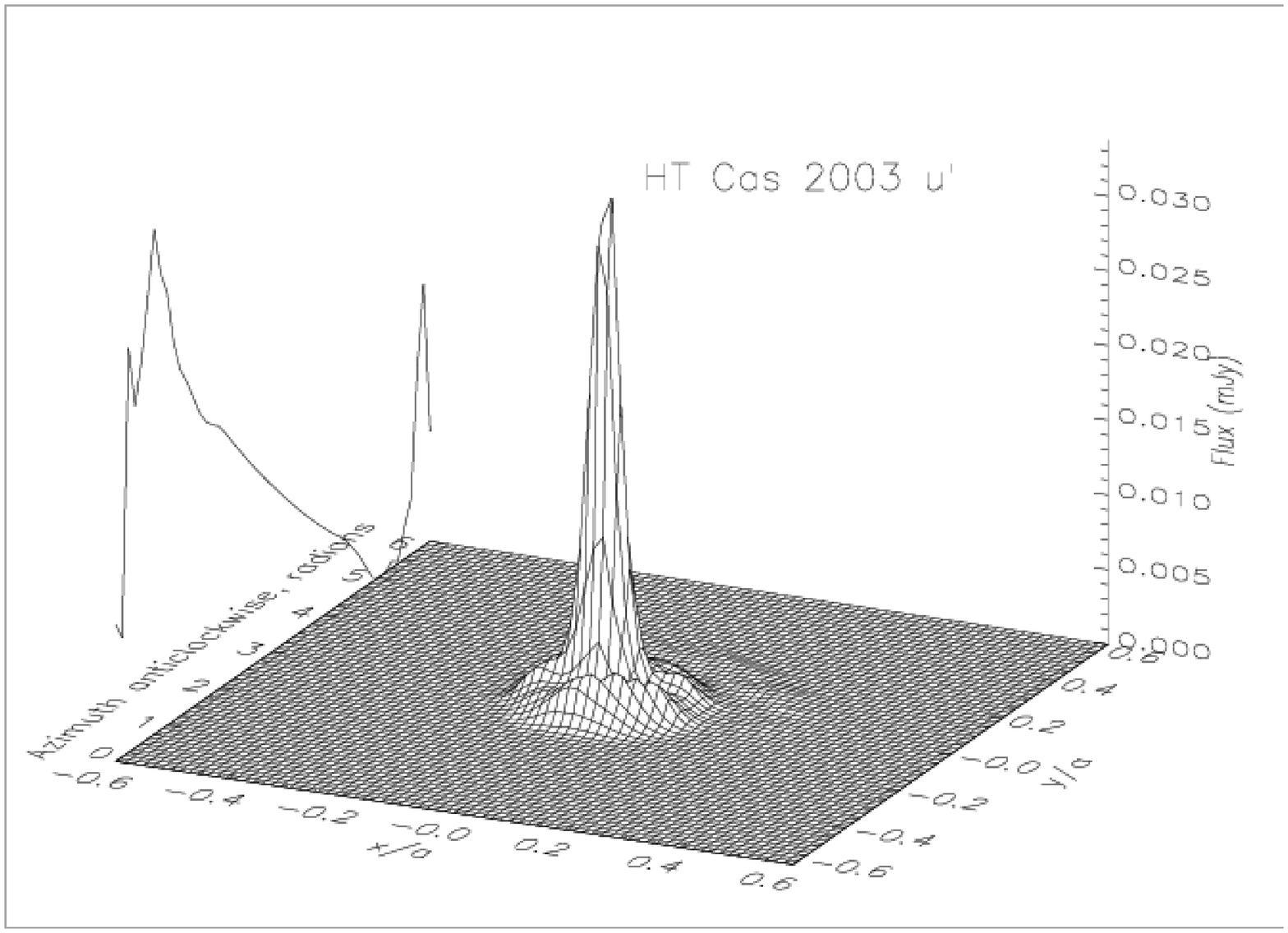,width=7.0cm,angle=0.} &
\psfig{figure=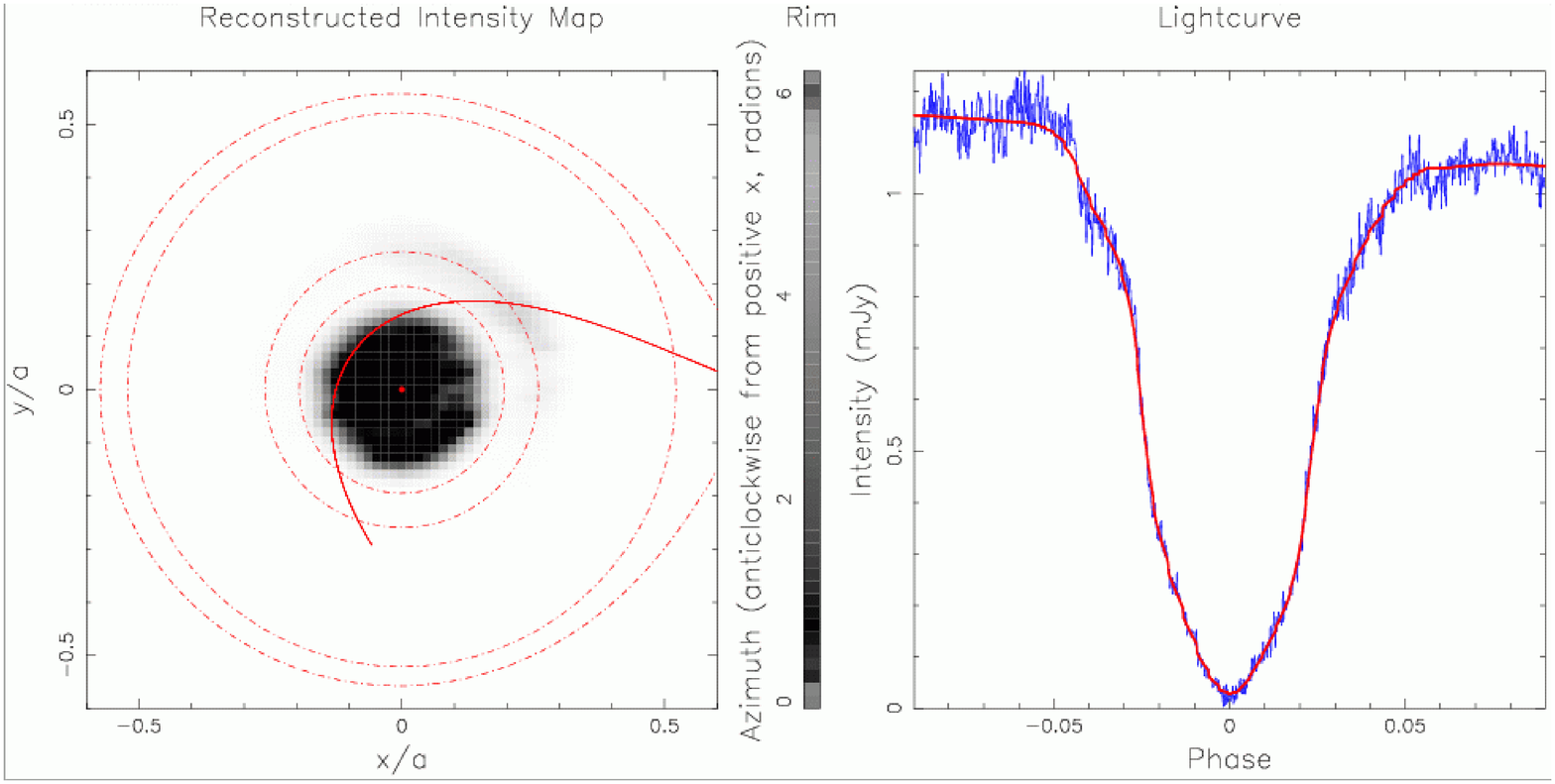,width=10.0cm,angle=0.} \\
\psfig{figure=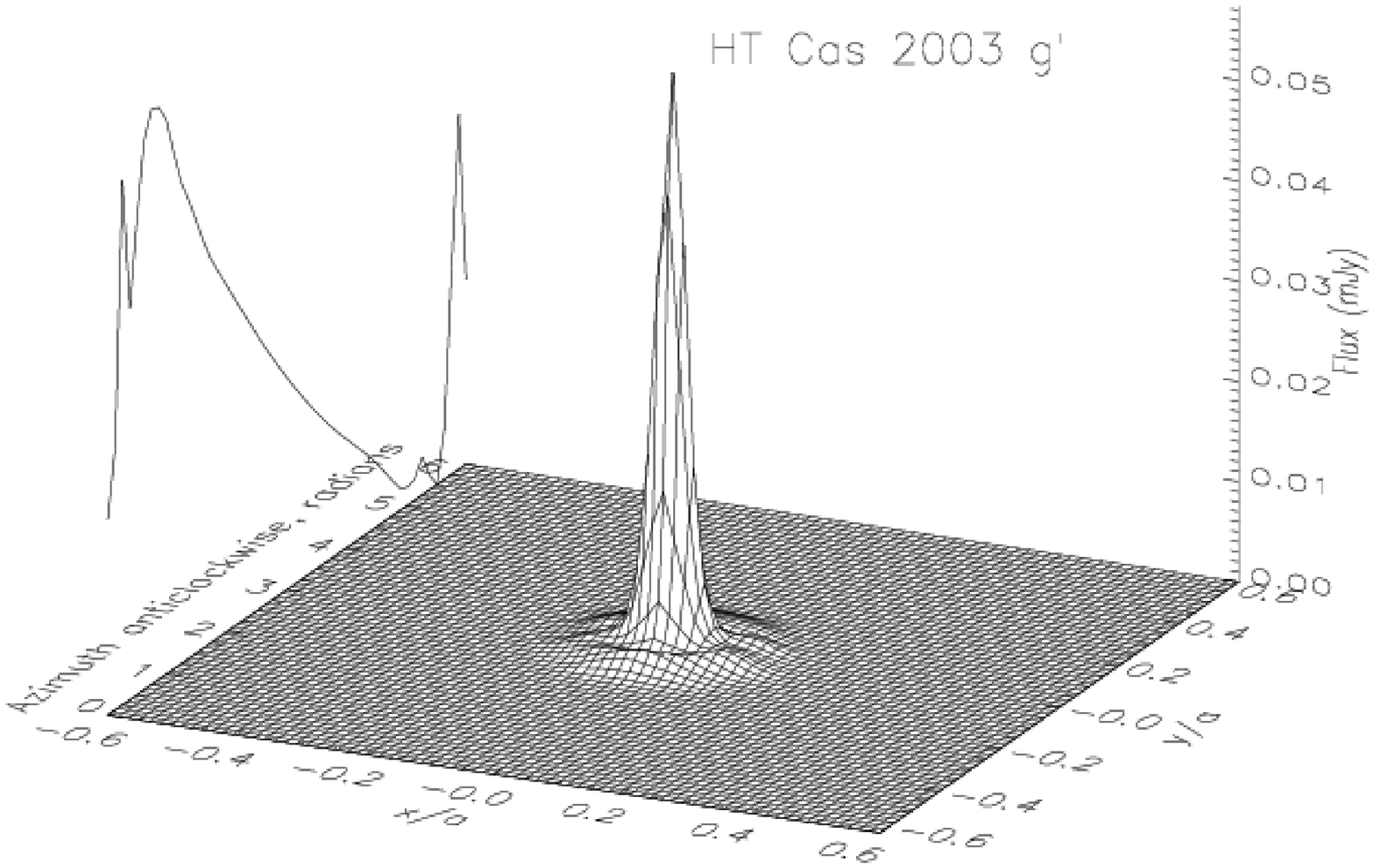,width=7.0cm,angle=0.} &
\psfig{figure=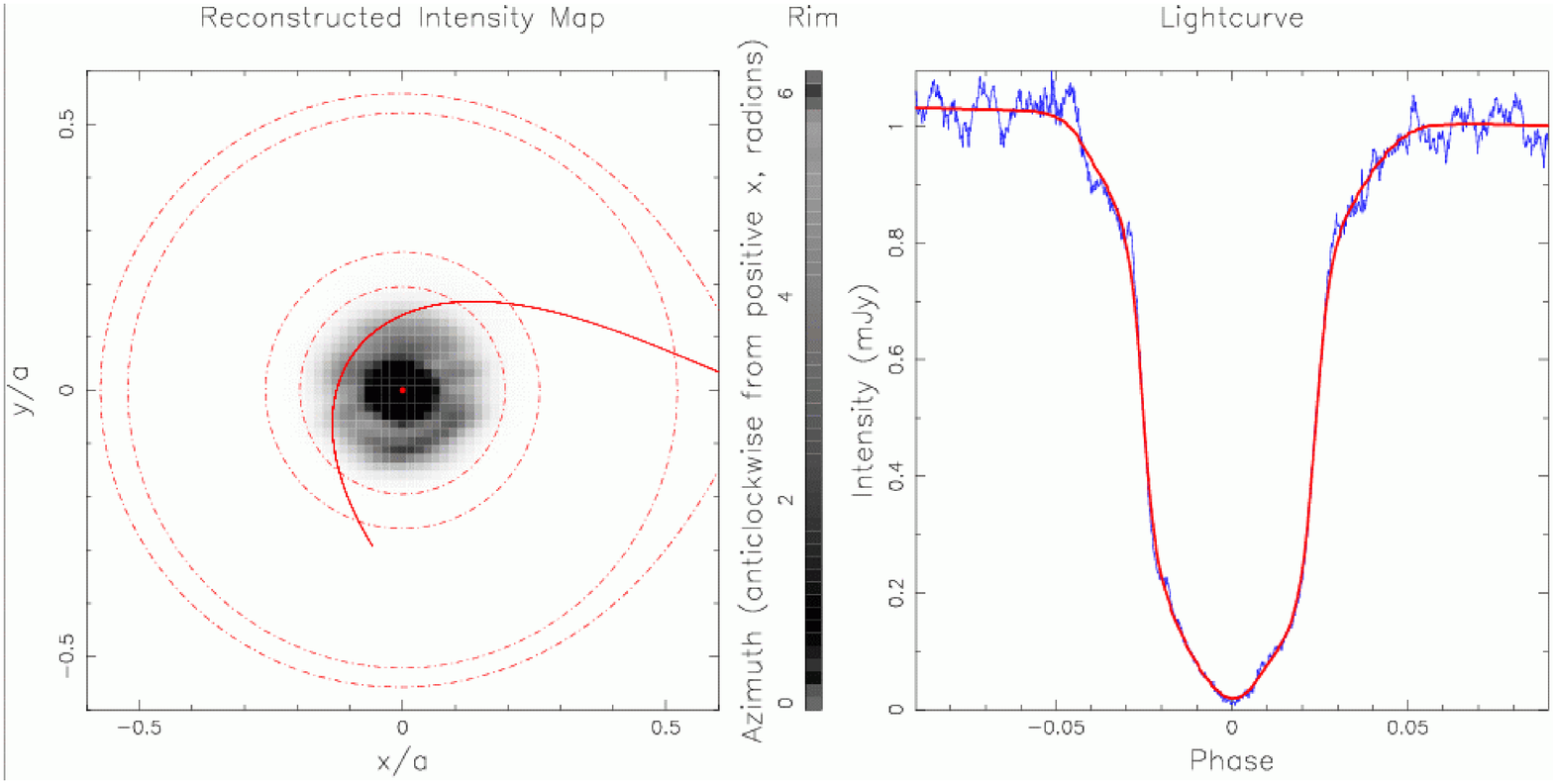,width=10.0cm,angle=0.} \\
\psfig{figure=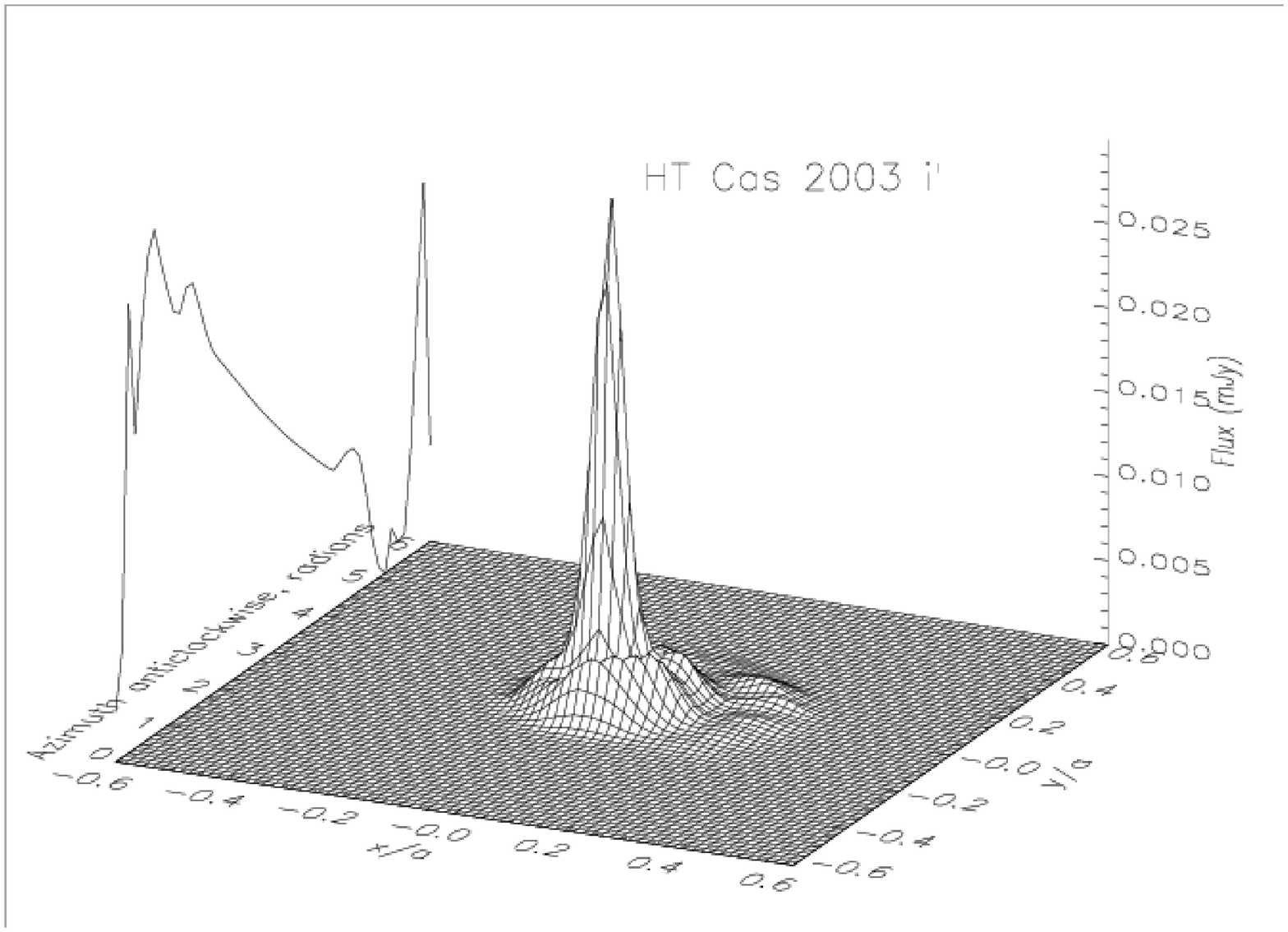,width=7.0cm,angle=0.} &
\psfig{figure=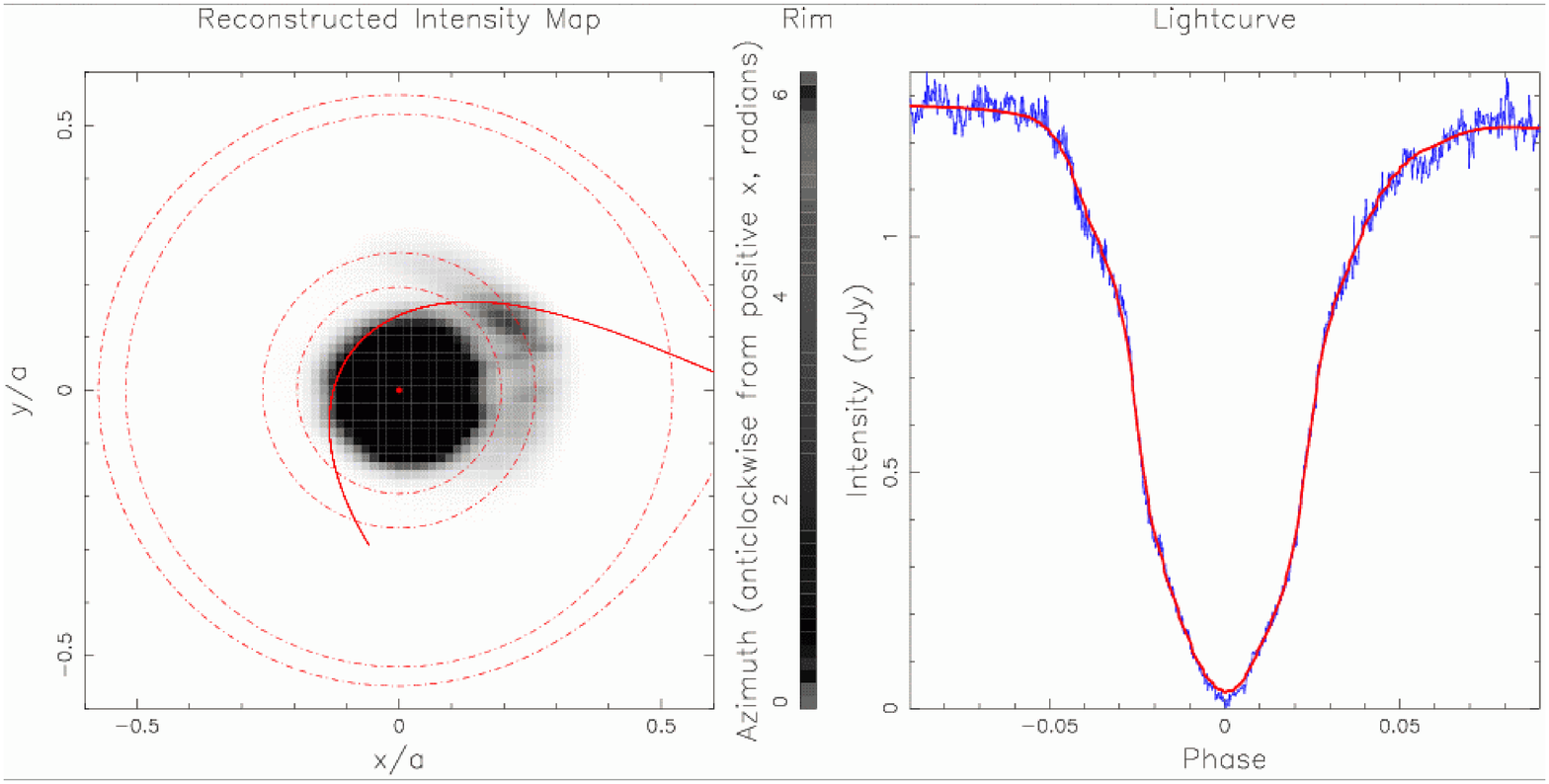,width=10.0cm,angle=0.} \\
\end{tabular}
\caption{As Fig.~\ref{fig:htcas_2002}, but for the {\em u}$^{\prime}$
  (top), {\em g}$^{\prime}$ (middle) and {\em i}$^{\prime}$ (bottom)
  data of 2003 October 29--30. The radius of the disc (the same as
  that of the disc rim) was estimated from the position of the bright
  spot in the eclipse maps, and is $0.26a$. The {\em u}$^{\prime}$,
  {\em g}$^{\prime}$ and {\em i}$^{\prime}$ light curves were offset
  vertically by 0.24, 0.14 and 0.42~mJy, respectively.}
\label{fig:htcas_2003}
\end{figure*}

\section{HT~Cas}
\label{sec:htcas}
\subsection{Light-curve morphology}

The light curves of HT~Cas shown in Fig.~\ref{fig:lightcurves} are
typical of those found in the literature \citep[e.g.][]{patterson81,
horne91b}. The typical out-of-eclipse {\em g}$^{\prime}$ flux for the
2002 data is 0.7~mJy (16.8~mag), and the typical mid-eclipse {\em
g}$^{\prime}$ flux is 0.1~mJy (18.9~mag; see also
Fig.~\ref{fig:htcas_2002}). The peak {\em g}$^{\prime}$ flux in the
2002 dataset is approximately 1.0~mJy (16.4~mag). HT~Cas is slightly
brighter in the 2003 data: the typical out-of-eclipse {\em
g}$^{\prime}$ flux is 1.2~mJy (16.2~mag), and the mid-eclipse {\em
g}$^{\prime}$ flux is again about 0.1~mJy (18.9~mag; see also
Fig.~\ref{fig:htcas_2003}). The 2003 data set is entirely consistent
with HT~Cas being in its high quiescent state \citep{robertson96}, but
the 2002 data appears to be somewhere between the high and low states
described therein. The data from 2002 show other clear differences
from the data of 2003. First, the eclipse bottoms in 2002 are much
flatter than those of 2003, implying that the brightness distribution
in 2002 was more centrally concentrated than in 2003. Second, the
eclipse depth in 2003 was greater than in 2002, which, as we shall see,
is due to increased disc emission and not an increase in the
brightness of the white dwarf. There is also visible in the 2002 data
a clear shoulder during egress. This appears to be the egress of the
bright spot, similar in appearance to the feature seen in the light
curves of \citet{patterson81}. Unfortunately for the aim of
determining the system parameters from the eclipse contact phases, no
bright spot ingress is visible. This variability of the light curve is
typical of HT~Cas \citep[e.g.][]{patterson81,horne91b,robertson96}.
The fact that the flux increase between 2002 and 2003 is slight
illustrates that these brightness and morphological variations are not
due to an overwhelming increase in the disc flux `drowning out' the
bright spot, as occurs during outburst. Observations of HT~Cas by the
AAVSO (Waagen, private communication; Fig.~\ref{fig:htcas_aavso}) show
no evidence for  an outburst of HT~Cas before the 2003 October
observations -- the change in the light curve is not due to the system
being on the way up or down from an outburst. We term the 2003 and
2002 data the `high' and `low' quiescent states, respectively
(although we caution that these probably differ from the various high
and low quiescent states discussed in the literature).


\subsection{Eclipse mapping}
\label{sec:eclipse_mapping}

The eclipse mapping method was developed by \citet{horne85}. It
reconstructs the two-dimensional brightness distribution of the
(assumed flat) accretion disc from the information contained in the
one-dimensional light curve. The disc is divided into a Cartesian
grid, centred on the white dwarf, each element of which has an equal
area. The intensity of each element is an independent parameter, which
is adjusted by an iterative procedure to find the best fit to the
light curve. The summation of the intensities of each visible grid
element at each phase produces the light curve. The eclipse geometry
is defined by the mass ratio $q$ and the orbital inclination $i$. The
parameters adopted for these reconstructions were those derived by
\citet{horne91b}, $q=0.15$ and $i=81\fdg0$. The fit is constrained by
minimising $\chi^{2}$ and maximising the entropy with respect to some
default map \citep{horne85}. This latter constraint is required since
the one-dimensional data  cannot fully constrain the two-dimensional
reconstruction of the disc intensity. One would expect Keplerian shear
to minimise any azimuthal structure in the accretion disc, so the
default map is usually chosen so that it suppresses azimuthal
structure in the disc whilst preserving the radial
structure. \citet{rutten93} used default maps of limited azimuthal
smearing, which attempt to preserve a degree of azimuthal structure in
the reconstructed intensity distribution. We have adopted a default
map of limited azimuthal smearing (constant angle;
\citealt*{baptista00}), with $\Delta\theta=0.7$~radians and $\Delta
R=0.01a$. We used the maximum entropy optimisation package {\sc
memsys} \citep{skilling84} to perform the iterative procedure. The
grid was a $75\times75$ pixel array; each element was further
subdivided into a $7\times7$ grid in order to determine its fractional
visibility at each orbital phase. This subdivision process saves
computational time during the iterative procedure whilst increasing
the quality of the final reconstruction. The 2002 {\em
u}$^{\prime}$ and {\em g}$^{\prime}$ light curves and all the 2003
light curves were rebinned by a factor of three (using a weighted
mean) in order to reduce flickering and to make them comparable to the
2002 {\em i}$^{\prime}$ data.

\begin{figure}
\begin{tabular}{c}
\psfig{figure=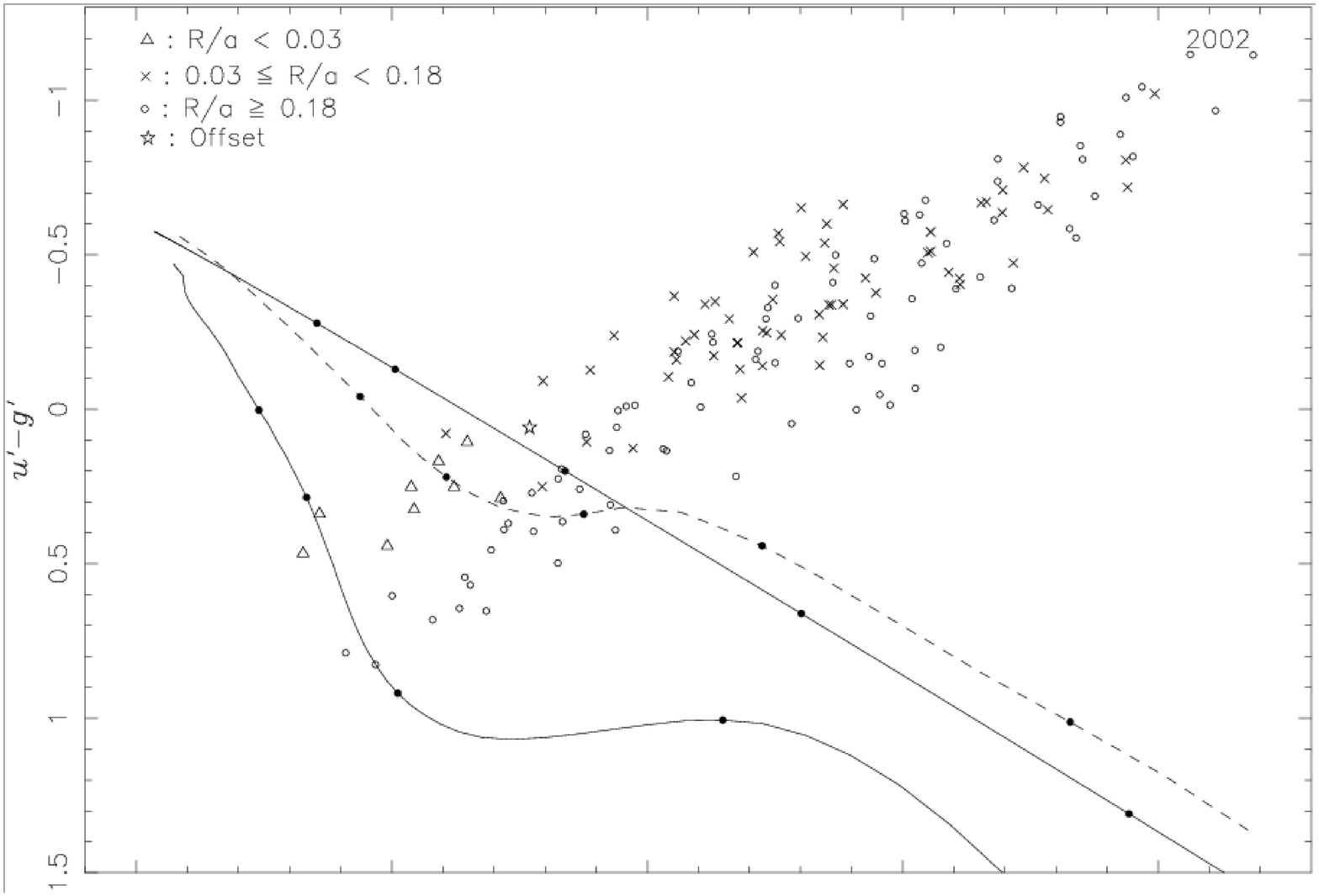,width=8.0cm,angle=0.} \\
\psfig{figure=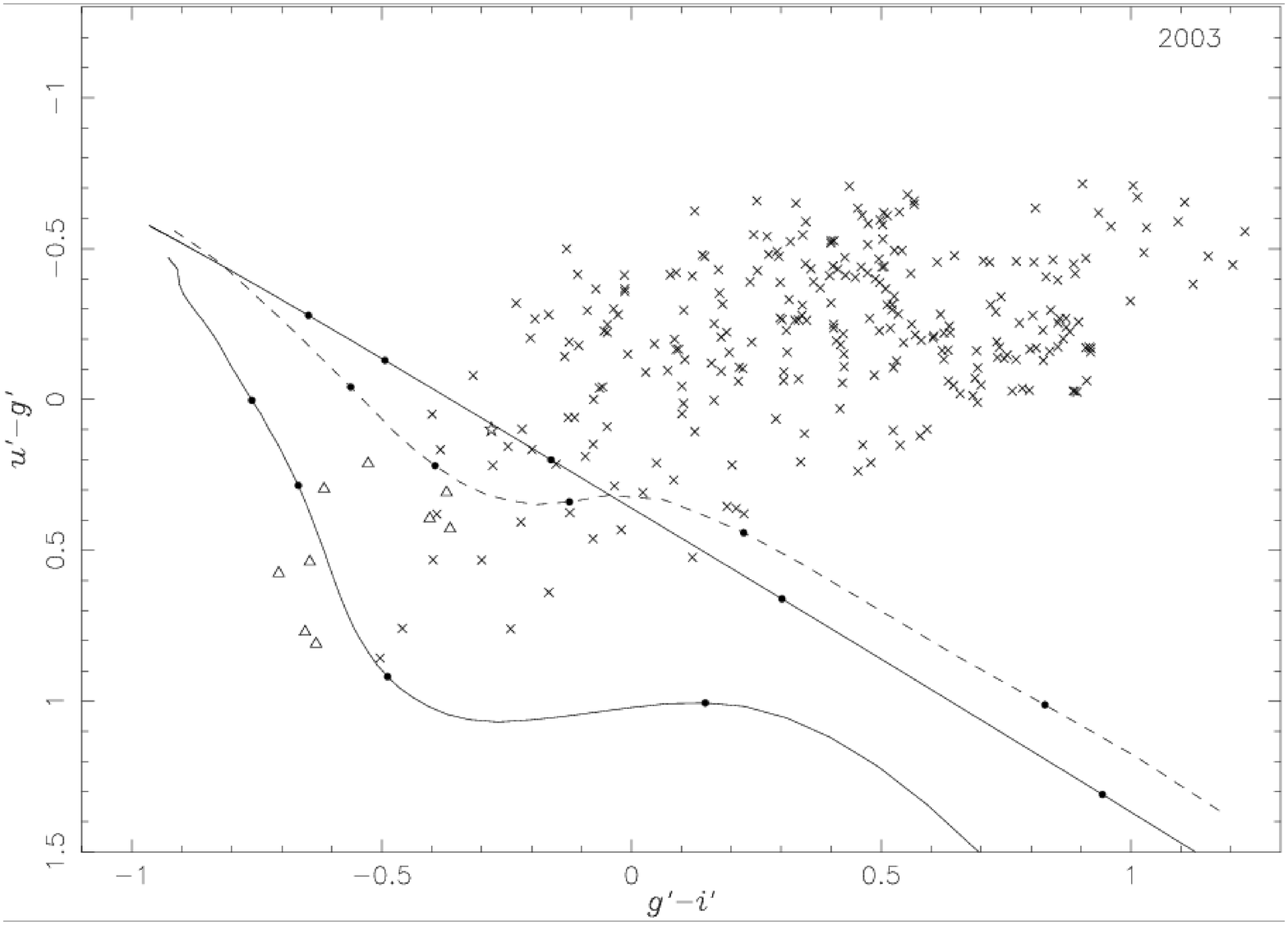,width=8.0cm,angle=0.}
\end{tabular}
\caption{Colour-colour diagrams of the accretion disc of HT~Cas in
  (top) 2002 and (bottom) 2003. The solid straight line is a blackbody
  relationship, the solid curve is the main-sequence relationship of
  \citet{girardi04} and the dashed curve is the white dwarf model
  atmosphere relation of \citet{bergeron95} described in
  section~\ref{sec:eclipse_mapping}. The filled circles superimposed
  upon each of these lines indicate temperatures of $20\,000$,
  $15\,000$, $10\,000$, $7000$ and $5000$~K, with the hotter
  temperatures located at the upper left of the plots (the $5000$~K
  point for the main-sequence curve lies off the plot). Each of the other
  points represents one element of the eclipse map. Elements at
  different radial distances $R$ from the centre of the white dwarf
  are plotted using different markers, as indicated in the figure. The
  position of the mid-eclipse (offset) flux is also plotted. In the
  interests of clarity, only points where the flux in all three
  passbands was greater than $5\times10^{-4}$~mJy were plotted.}
\label{fig:htcas_colours}
\end{figure}

As the eclipse mapping method assumes that all the light originates
from the accretion disc, any light remaining uneclipsed at mid-eclipse
breaks the anti-correlation between the relative eclipse depth and the
eclipse width implied by this assumption. This additional light is
placed by the maximum entropy
procedure in those parts of the disc least constrained by the data,
such as the region farthest from the secondary star (the `back') of
the disc. Therefore, a constant was subtracted from the light curves
prior to fitting. Ideally, this offset would be estimated by maximum
entropy methods, either by computing a series of eclipse maps with
differing offset values and selecting the map with the highest entropy
or by including the offset as an additional free parameter in the
eclipse mapping code
\citep{rutten92b,rutten94a,baptista95,baptista96}. Unfortunately, these
techniques both fail for highly asymmetric accretion discs
\citep{baptista96} such as that of HT~Cas in quiescence, because the
spurious structure introduced in the reconstructed map by the
uneclipsed component mixes with the asymmetric emission, forming a
more symmetrical structure in the disc. This increases the entropy of
the reconstructed map, meaning that the map with the largest entropy
is not that with the correct offset. The offsets for these data were
therefore first estimated from the mid-eclipse flux level, and
fine-tuned by computing a series of eclipse maps with different offset
values and selecting the map with the least spurious structure. These
offsets are given in the relevant figure captions and are discussed
further in section~\ref{sec:discussion} and Fig.~\ref{fig:htcas_colours}.

The maximum entropy method assumes that all variations in the light
curve are due to the eclipse by the secondary star. As such, the
orbital hump, which is due to the changing foreshortening of the
bright spot, can introduce spurious features into the reconstructed
map of the disc intensities. For example, \citet{baptista00}
attempted to correct for the orbital hump by fitting a spline function
to phases outside eclipse, dividing the light curve by the fitted
spline and scaling the result to the value of the spline function at
phase zero \citep[see also][]{horne85}. This technique worked well on
the light curve of IP~Peg in outburst to which \citet{baptista00}
applied it, but it is not suitable for the light curves of quiescent
dwarf novae because not all parts of the disc contribute to the
orbital hump: the white dwarf emits isotropically whereas the bright
spot does not \citep{horne85,bobinger97}. The orbital hump was
accounted for by a disc rim \citep[e.g.][]{bobinger97}, divided into
50 segments of equal angle. The rim was assumed to be of negligible
height, so that the flat disc assumption is not violated. We have
estimated the radius of the disc rim from the position of the
reconstructed bright spot, since the disc radius, of $0.28a$ in 2002
and $0.26a$ in 2003, is larger than that derived by \citet{horne91b}
of $0.23a$.

We have chosen not to deconvolve and remove the white dwarf from the
light curves as the presence of flickering and the lack of a clear
distinction between the eclipses of the white dwarf, bright spot and
accretion disc makes this difficult to do so reliably (see, for
example, \citetalias{feline04a}). Besides, clear evidence for the
features reproduced in the eclipse maps and discussed below can be
seen directly in the light curves themselves.

As only a few light curves were used, the noise in the light curves is
dominated by flickering rather than by photon noise. Iterating to a
reduced $\chi^{2}=1$ is therefore inappropriate in this case and leads
to the noise in the light curves (flickering) being transposed to the
eclipse maps. Consequently, the eclipse maps were computed by
progressively relaxing the $\chi^{2}$ constraint until the noise in
the eclipse maps was satisfactorily ameliorated (as judged by visual
inspection).

\begin{figure}
\begin{tabular}{c}
\psfig{figure=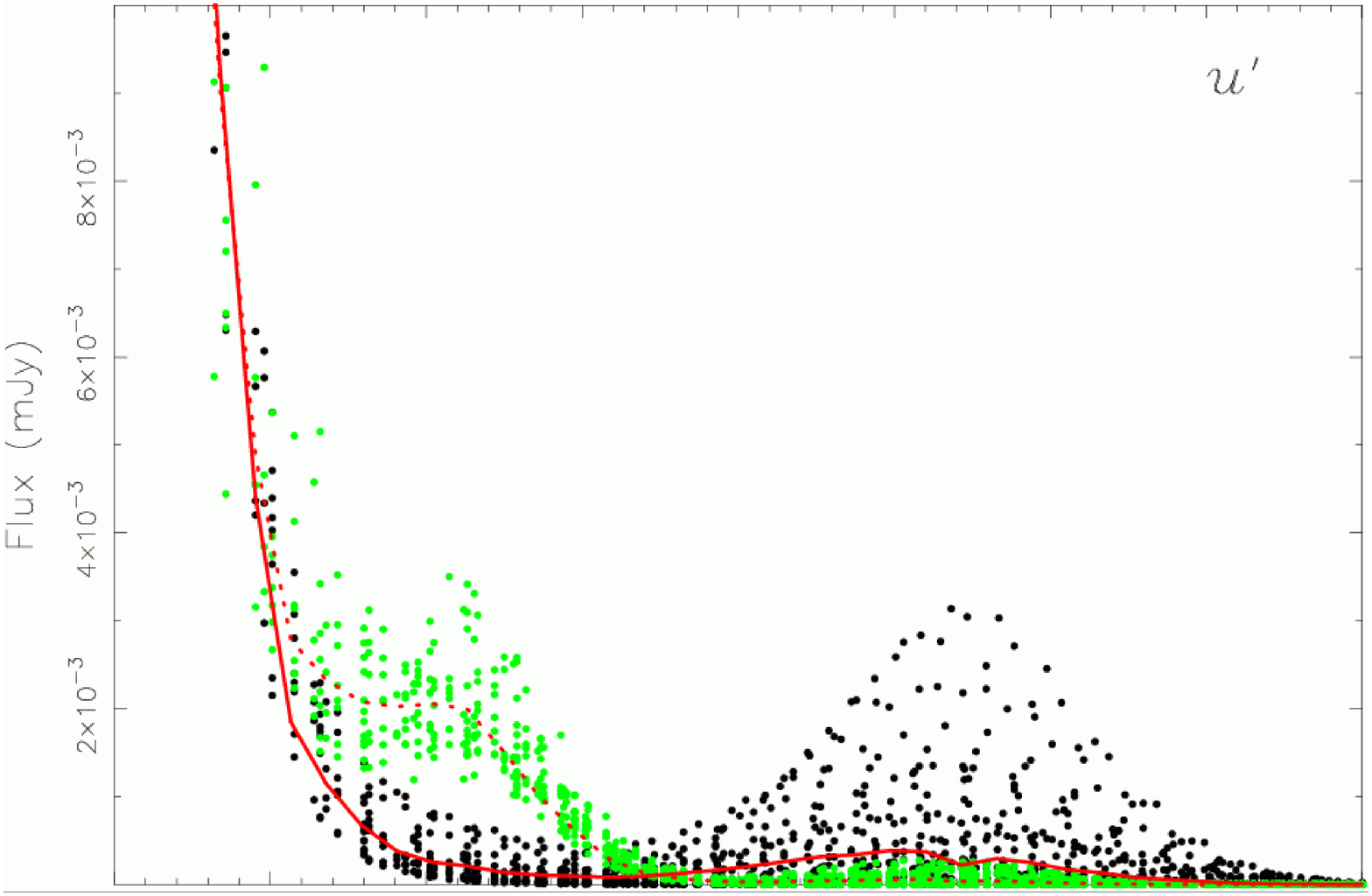,width=8.0cm,angle=0.} \\
\psfig{figure=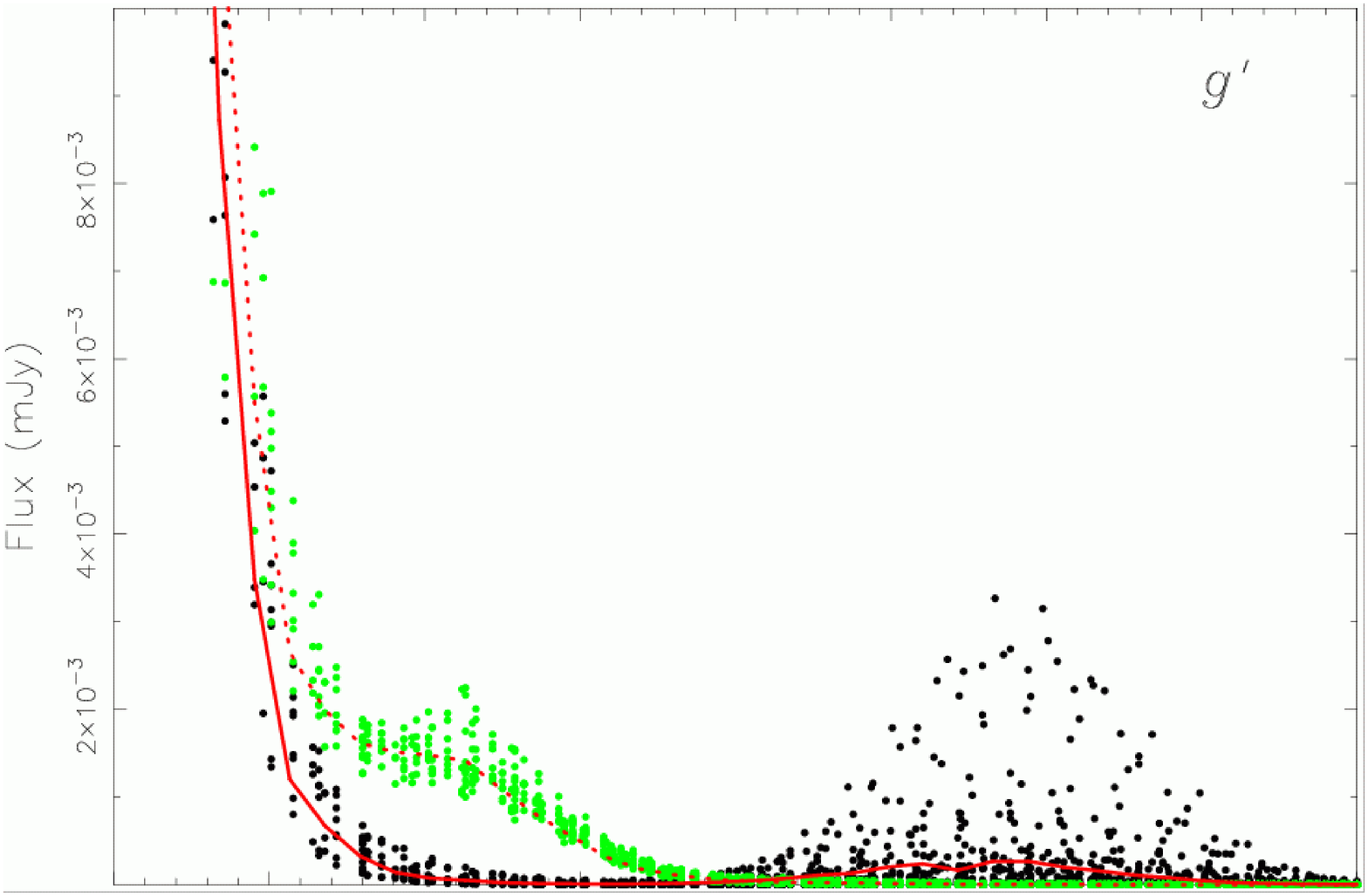,width=8.0cm,angle=0.} \\
\psfig{figure=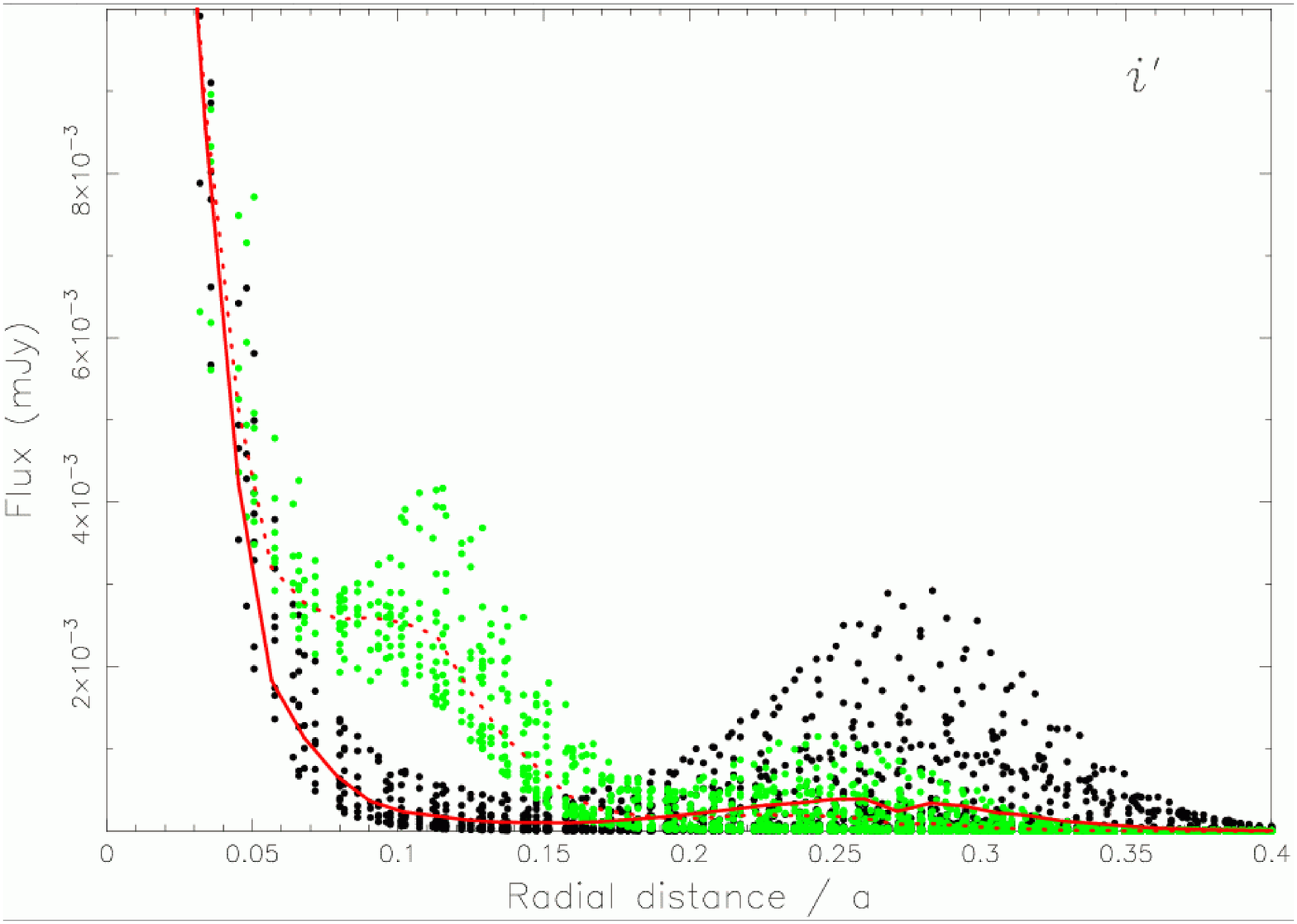,width=8.0cm,angle=0.} \\
\end{tabular}
\caption{The radial flux distribution of the reconstructed accretion
  disc of HT~Cas for the 2002 data (black dots and solid red line) and
  the 2003 data (green dots and dotted red line). The dots represent
  the flux and radius of the individual grid elements; the red lines
  represent the mean flux in concentric annuli. The {\em u}$^{\prime}$
  and {\em g}$^{\prime}$ flux distributions were determined using the
  data of 2002 September 13--14 and 2003 October 29--30, whereas the
  {\em i}$^{\prime}$ distributions were determined using the data of
  2002 September 13 and 2003 October 29--30 (due to a loss of
  sensitivity in the {\em i}$^{\prime}$ chip on 2002 September 14.)}
\label{fig:htcas_radial}
\end{figure}

The reconstructed eclipse maps of HT~Cas shown in
Figs~\ref{fig:htcas_2002} and \ref{fig:htcas_2003} show distinct
morphological changes from 2002 to 2003. In both 2002 and 2003 the
flux comes primarily from the white dwarf, but in 2002 there is clear
evidence for a faint bright spot in the outer regions of the accretion
disc. The 2003 eclipse maps show a very weak bright spot in the {\em
u}$^{\prime}$ and {\em i}$^{\prime}$ passbands only. The absence of a
bright spot in the 2003 {\em g}$^{\prime}$ band may, however, be a
contrast effect, as the white dwarf is approximately twice as bright
in {\em g}$^{\prime}$ as in the other bands. Not only is the bright
spot much fainter/absent in the 2003 reconstructed maps, but there is
evidence for emission from the inner portions of the accretion
disc. This is best illustrated by the radial flux profiles shown in
Fig.~\ref{fig:htcas_radial}. Comparing the two radial profiles to the
light curves shown in Fig.~\ref{fig:lightcurves} demonstrates that the
increased emission from the inner disc corresponds to a higher
brightness state. The greater flux in 2003 is not, upon inspection of
the eclipse maps, due to increased emission from the white dwarf,
which is actually fainter in 2003 than in 2002, but due to a brighter
inner disc.

By summing the flux from each element of each eclipse map whose centre
lies within $0.03a$ of the centre of the white dwarf and fitting the
resulting colours to the hydrogen-rich, $\log g=8$ white dwarf model
atmospheres of \citet{bergeron95}, converted to the SDSS system using
the observed transformations of \citet{smith02}, the temperature of
the white dwarf was determined to be $T_{\rm w}=15\,000\pm1000$~K in
2002 and $T_{\rm w}=14\,000\pm1000$~K in 2003. Varying the distance
from the white dwarf over which the summation took place between
$0.01a$--$0.07a$ did not significantly affect the colours and
therefore did not significantly affect these temperature
estimates. These temperatures are consistent with those found by
\citet[$T_{\rm w}=14\,000\pm1000$~K]{wood92} and by \citet[$T_{\rm
w}=15\,500$~K]{vrielmann02} from the same set of quiescent photometric
observations. The effect of the variable nature of the accretion disc
of HT~Cas is evident in the white dwarf temperatures of HT~Cas derived
by \citet{wood95a}, of $T_{\rm w}=13\,200\pm1200$~K during a low state
(which is consistent with our results) and $T_{\rm
w}=18\,700\pm1800$~K during a normal state (which differs from our
results by $\sim2\sigma$). 

In Fig.~\ref{fig:htcas_colours} we present the colour-colour diagrams
for the 2002 and 2003 eclipse maps of HT~Cas. In both 2002 and 2003
the scatter of the data points from the central regions of the disc
($R/a<0.03$), comprising the white dwarf and boundary layer, is
consistent with an increase in the {\em g}$^{\prime}$ flux over that
expected from a lone white dwarf shifting the position of the
datapoint down and to the left on Fig.~\ref{fig:htcas_colours}. This
suggests a contribution to this flux from the boundary layer
surrounding the white dwarf. This excess {\em g}$^{\prime}$ flux is
(marginally) more pronounced in 2003 than in 2002, as might be
expected given the differences between the distribution of the disc
flux for these dates. The emission from both the inner ($0.03\leq
R/a<0.18$) and outer ($R/a\geq0.18$) regions of the disc are
concentrated to the right of the blackbody relation in
Fig.~\ref{fig:htcas_colours}, suggesting that the disc is optically
thin \citep{horne85a,wood92,baptista96}. There seems to be no
significant difference between the colours of the inner and outer
discs of 2002 (in 2003 the outer disc was too faint to be plotted on
Fig.~\ref{fig:htcas_colours}). Interestingly, in both 2002 and 2003,
the offset colours lie on the blackbody relation rather than being on
the main-sequence curve. A blackbody fit to the 2002 and 2003 offset
colours gave $T\sim11\,000$~K in each case.

\section{Discussion}
\label{sec:discussion}

\defcitealias{kato02a}{Kato et al.\ 2002a}

We have found that the dwarf novae GY~Cnc and IR~Com both exhibit
eclipses of the white dwarf, and have a bright spot which is faint
(GY~Cnc) or undetected (IR~Com). We have determined updated
ephemerides for both of these objects. IR~Com, with its short orbital
period, significant flickering, high/low quiescent states
\citep{richter95,richter97} and lack of orbital hump or bright spot
strongly resembles HT~Cas in terms of its photometric behaviour
\citepalias[see also][]{kato02a}.

The colours of the offset flux of HT~Cas shown in
Fig.~\ref{fig:htcas_colours}, which is estimated from the flux at
mid-eclipse, suggest that it does not originate solely from the
secondary star. \citet{marsh90c} detected the secondary star in
HT~Cas, and found it to be indistinguishable from a main-sequence star
of spectral type M$5.4\pm0.25$, which lies off to the bottom right of
the plot of Fig.~\ref{fig:htcas_colours} on the main-sequence
relation. We think it unlikely that the mid-eclipse flux is from outer
regions of the accretion disc at the back of the disc which remain
uneclipsed at phase zero, since examination of the eclipse maps shown
in Figs~\ref{fig:htcas_2002} and \ref{fig:htcas_2003} reveals that the
emission from the rest of the disc is restricted to either the bright
spot (in 2002) or the inner disc (in 2003). Given this, our preferred
explanation for the mid-eclipse colours of HT~Cas is that they are a
combination of flux from the secondary star (which dominates in the
{\em i}$^{\prime}$ band) and flux from a vertically extended,
optically thin disc wind, whose Balmer emission causes it to dominate
in the {\em u}$^{\prime}$ band. We note that the offset flux level
seems to be correlated with the flux from the inner regions of the
accretion disc, which supports the hypothesis of a disc wind
originating from the inner disc region or boundary layer of HT~Cas. We
caution, however, that systematic errors may affect the offset fluxes
due to the technique used to determine them, and that this conclusion
is therefore tentative. (As the offset flux is a small fraction of the
total light, except in the {\em i}$^{\prime}$ band, any systematic errors
present in the offset fluxes will not significantly affect the rest of
our results.)

The eclipse maps of HT~Cas shown in Figs~\ref{fig:htcas_2002} and
\ref{fig:htcas_2003} and the radial flux profiles shown in
Fig.~\ref{fig:htcas_radial} clearly demonstrate that the accretion
disc of HT~Cas was in two distinct states during our 2002 and 2003
observations. In the 2002 data the disc provided a negligible
contribution to the total light, except for the presence of a bright
spot in its outer regions. In 2003 the bright spot was much fainter,
but the inner disc was luminous, causing the overall system brightness
to be $\sim0.6$~mJy brighter than in 2002. The uneclipsed component
was also slightly brighter in 2003 than 2002 (see captions to
Figs~\ref{fig:htcas_2002} and \ref{fig:htcas_2003}), but was not the
major cause of the differences in the flux. We proceed to review
previous observations and to discuss various possible explanations for
this behaviour.

The most likely reasons for the observed changes in the intensity
distribution of the quiescent accretion disc of HT~Cas lie in
variability of the secondary star (which supplies the disc with
material) or some property of the accretion disc itself. We can
exclude the white dwarf as the cause of the variability since the only
plausible way that this could affect the outer regions of the disc is
via a magnetic field. HT~Cas is a confirmed dwarf nova, whose white
dwarfs do not have magnetic fields strong enough to significantly
influence the motion of gas in the disc \citep[e.g.][]{warner95}.

The most obvious mechanism for the accretion disc to produce the
observed behaviour of HT~Cas is via some relationship to the outburst
cycle. \citet{baptista01} reported changes in the structure of the
accretion disc of EX~Dra (a dwarf nova above the period gap) through
its outburst cycle, specifically the presence of a low-brightness
state immediately after outburst during which the disc and bright spot
were exceptionally faint. In EX~Dra, the low-brightness state is due
to reduced emission from all parts of the disc and white dwarf; our
results, however, demonstrate that the quiescent luminosity of HT~Cas
depends on which areas of the disc are luminous. \citet{robertson96}
find that both the transition between the quiescent high and low
states and the duration of the low state in HT~Cas occur on
time-scales of days to months compared to the outburst cycle length of
$\sim400$~days \citep{wenzel87}. \citet*{truss04} proposed a (slowly
cooling) hot inner region of the disc in order to explain the constant
quiescent brightness observed in (most) dwarf novae, which is contrary
to the increase of 1--3 magnitudes predicted by most disc instability
models (see \citealp{lasota01} for a review). This fails to account
for the observed changes in the outer accretion disc of HT~Cas, but does
provide a plausible explanation for the variability of the inner
regions of the disc. This model, however, necessitates an outburst
between the two sets of observations reported here, which amateur
observations (Fig.~\ref{fig:htcas_aavso}) can neither confirm nor
refute. We conclude that this latter scenario is the only plausible
way in which the changes in the accretion disc of HT~Cas could be
related to its position in the outburst cycle.

The secondary star can also induce changes in the accretion disc. For
example, variability of the rate of mass transfer from the secondary
star is often (plausibly) cited as a mechanism to explain the
quiescent variability of dwarf novae (and other
CVs). \citet{baptista04} observed the short-period dwarf nova
V2051~Oph in high and low quiescent states. Eclipse maps showed that
the increased emission in the high state was due to greater emission
from the bright spot and gas stream region, implying a higher mass
transfer rate from the secondary star. Interestingly, this is the
opposite to what we find for HT~Cas.

Variability of the secondary star is in fact the mechanism usually
proposed to explain the well-documented presence of high/low
quiescence states in HT~Cas (e.g.\ \citealp*{berriman87},
\citealp{wood95,robertson96}). The most frequently cited explanation
is that suggested by \citet{livio94}, of star spots passing over the
inner Lagrangian point temporarily lowering the mass transfer rate
from the secondary star. Another possible causal process is magnetic
variability of the secondary star. \citet*{ak01} found cyclical
variations in the quiescent magnitudes and outburst intervals of 22
CVs, which they attributed to solar-type magnetic activity cycles of
the secondary stars. This can result in an increased mass transfer
rate from the secondary star as well as the removal of angular
momentum from the outer regions of the disc, causing material to
accumulate in the inner regions of the disc rather than in the outer
regions. The magnetic activity cycle of the secondary stars derived by
\citet{ak01} is, however, on the wrong time-scale (years) to explain
the frequency of the high/low state transitions and durations
(days/months) observed by \citet{robertson96}.

In summary, variations in the rate of mass transfer from the secondary
star can explain the variability of the bright spot, but fail to
account for the changes in the inner disc. These can be explained by a
larger mass transfer rate through the accretion disc (possibly due to
a rise in the disc viscosity and/or the scenario proposed by
\citealt{truss04}) increasing the emission from the inner disc via
viscous dissipation.

We conclude that the variability of the quiescent accretion disc of
HT~Cas is caused by variations both in the rate of mass transfer from
the secondary star and through the accretion disc. In our 2002
observations then, the rate of mass transfer through the disc was
lower and the rate of mass transfer from the secondary star greater
than in 2003. It is clearly desirable to undertake long-term
monitoring of HT~Cas (or a similar object, e.g.\ IR~Com) with the aim
of eclipse mapping the changes that occur in the disc during
quiescence and especially during a transition between the high and low
states in order to determine the triggers and physical mechanisms
underlying this behaviour.


\section*{Acknowledgments}
We thank the anonymous referee for useful comments.
WJF is supported by a PPARC studentship. TRM acknowledges the support
of a PPARC Senior Research Fellowship. ULTRACAM is supported by PPARC
grant PPA/G/S/2002/00092. This research has made use of NASA's
Astrophysics Data System Bibliographic Services. Based on observations
made with the William Herschel Telescope operated on the island of La
Palma by the Isaac Newton Group in the Spanish Observatorio del Roque
de los Muchachos of the Instituto de Astrofisica de Canarias. We
acknowledge with thanks the variable star observations from the AAVSO
International Database contributed by observers worldwide and used in
this research.

\bibliographystyle{mn2e} 
\bibliography{/home/wf/paper/ouvir/abbrev,/home/wf/paper/ouvir/refs}

\end{document}